\documentclass{article}
\usepackage[utf8]{inputenc}
\usepackage[colorinlistoftodos]{todonotes}
\usepackage{placeins}
\usepackage{amsmath}
\usepackage{subfig}
\usepackage{authblk}

\usepackage{mathtools}
\makeatletter
\newcommand{\Biggg}{\bBigg@{4}}
\newcommand{\Bigggg}{\bBigg@{5}}
\makeatother

\title{Resonance driving terms and tune spread with amplitude driven a beam-beam long range kick and a DC wire kick}
\author[1]{K.~Skoufaris}
\author[1]{G.~Sterbini}
\author[1]{Y.~Papaphilippou }
\affil[1]{European Organization for Nuclear Research (CERN), CH-1211 Geneva 23, Switzerland}

\begin{document}

\maketitle

\begin{abstract}
This document presents in detail the derivation of the formulas that describe the resonance driving terms and the tune spread with amplitude generated by the beam-beam long range interactions and the DC wire compensators in cyclical machines. This analysis make use of the weak-strong approximation.
\end{abstract}

\begingroup
\color{black}
\tableofcontents
\endgroup

\pagebreak

\section{Introduction}
This document presents in detail the derivation of the formulas that describe the resonance driving terms (RDT) and the tune spread with amplitude (TSA) generated by the beam-beam long range (BBLR) interactions and the DC wire compensators in cyclical machines. For the study of the long range beam-beam interactions, the weak-strong approximation~\cite{baer,hira} is used. According to this, the particles in the ``weak'' beam interact with the electromagnetic field generated by the charge distribution of the ``strong'' beam while the latter beam is not affected by the charge distribution of the former one. For the calculation of the resonance driving terms (RDT), the treatment presented in~\cite{guig,beng} is followed while for the calculation of the tune spread with amplitude (TSA), the first order perturbation theory~\cite{gold} is used. 

\section{BBLR and CD wire compensator effective electric and magnetic fields}
The charge distributions that generate the BBLR kicks is described by a two dimensional Gaussian at the transverse plane and at the longitudinal one by a line distribution since $\sigma_s>>\sigma_\chi$ with $\chi=x,y$. From such a charge distribution, the resultant transverse electromagnetic field satisfies the formula $B_\phi=-\beta_{ls}E_r/c$ in the lab rest frame. Therefore, the integrated Lorentz force experienced by a test particle in the weak beam with negligible transverse velocity is given by: 
\begin{eqnarray}
	\label{lorfor}
	\int \vec{F}_{we}~ds=&&\int q(\vec{E}+\vec{u}_{we}\times \vec{B})~ds\Rightarrow \nonumber\\
	\int F_{we}~ds=&&q (1+\beta_{we}\beta_{ls})~\delta_D(s-s_0+u_{we}t) \int E_r~ds \nonumber\\
	=&&q (1+\beta_{we}\beta_{ls})~\frac{\delta_D(s-s_0)}{2} \int E_r~ds \nonumber\\
	=&&q(\mathcal{E}_{x}+\mathcal{E}_{y}).
\end{eqnarray}
$q$ is the electric charge of the test particle, $\delta_D$ is the Dirac delta function, $s=s_0+u_{we}t$, $\delta_D(2x)\equiv\delta_D(x)/2$,  $\beta_{ls}=-u_{st}/c$, $\beta_{we}=u_{we}/c$ and $c$ is the speed of light. The velocities $u_{st}$ and $u_{we}$ are measured in the lab rest frame and are the ones of the strong and weak bunches, respectively. A detailed derivation of the electromagnetic field that describe the beam-beam interactions is presented in~\cite{baer,hira}. Using these results the expressions for the effective electric ($\mathcal{E}_{\chi}$) and effective magnetic ($\mathcal{B}_{\chi}$) fields that describe the BBLR interactions for round ($\sigma_\chi=\sigma_\psi=\sigma$) and elliptical bunches ($\sigma_\chi>\sigma_\psi$) are given by:
\begin{subequations}
	\begin{equation}
		\mathcal{B}_{x}=\frac{\beta_{ls}}{c}\mathcal{E}_{y}
	\end{equation}
	\begin{equation}
		\mathcal{B}_{y}=-\frac{\beta_{ls}}{c}\mathcal{E}_{x}
	\end{equation}
	\begin{equation}
		\mathcal{E}=\mathcal{E}_{\psi}+\mathrm{i} \mathcal{E}_{\chi }=\mathcal{L} f \delta_D(s-s_0)			
	\end{equation}
	\begin{equation}
		\mathcal{L}=
		\begin{cases}
			\frac{N_p q (1+\beta_{we} \beta_{ls})}{4 \epsilon_0 \pi } & \ \ \ \text{for }\sigma_\chi=\sigma_\psi=\sigma \\
			
			\frac{N_p q (1+\beta_{we} \beta_{ls})}{4 \epsilon_0 \sqrt{\pi}\varDelta_\chi} & \ \ \ \text{for }\sigma_\chi>\sigma_\psi .
		\end{cases}
	\label{field_L}
	\end{equation}
	\begin{equation}
		f=
		\begin{cases}
			\frac{\psi+\widetilde{\psi}+\mathrm{i}(\chi+\widetilde{\chi}) }{(\chi+\widetilde{\chi})^2+(\psi+\widetilde{\psi})^2}\left( 1-\mathtt{Exp}\left[ -\frac{(\chi+\widetilde{\chi})^2+(\psi+\widetilde{\psi})^2}{2\sigma^2}\right] \right) & \ \ \ \text{for }\sigma_\chi=\sigma_\psi=\sigma \\
			
			\mathtt{Exp}\left[-\left( \frac{\chi+\widetilde{\chi}+\mathrm{i} (\psi+\widetilde{\psi})}{\varDelta_\chi} \right)^2 \right] \times \\
			\left( \mathtt{Erf}\left[ \frac{(\psi+\widetilde{\psi}) \sigma_\chi^2 - \mathrm{i} (\chi+\widetilde{\chi}) \sigma_\psi^2}{\sigma_\chi \sigma_\psi \varDelta_\chi}\right] + \mathtt{Erf}\left[ \frac{\mathrm{i} (\chi+\widetilde{\chi})-(\psi+\widetilde{\psi})}{\varDelta_\chi}\right] \right) & \ \ \ \text{for }\sigma_\chi>\sigma_\psi .
		\end{cases}
	\label{field_f}
	\end{equation}
	\label{bblr_em_eq}
\end{subequations}
For round bunches $(\sigma_\chi=\sigma_\psi=\sigma)$ $\chi=x$ and $\psi=y$ while for elliptical ones $(\sigma_\chi>\sigma_\psi)$ $\chi=x,y$ and $\psi=x$ if $\chi=y$ or $\psi=y$ if $\chi=x$. The symbols $(\widetilde{\chi},\widetilde{\psi})$ represent the transverse position of the weak bunch measured from the center of the strong bunch, where $(\chi,\psi)$ is the transverse position of the test particle measured from the center of the weak bunch, $\epsilon_0$ is the vacuum permittivity, $\varDelta_\chi=\sqrt{2\left( \sigma_\chi^2-\sigma_\psi^2\right) }$ and $\mathtt{Erf}\left[ \Xi \right] $ is the error function with $\Xi$ a complex number.

The magnetic field ($B_w$) generated by a wire compensator of length $L_w$ can be calculated with the use of the Biot-Savart law~\cite{jack}. However, since the wire length (a few meters long) is quite larger than the distance of the weak beam from the wire (less than a few centimeters), the magnetic field of an infinite long wire can be used. Thus, the integrated-effective magnetic field $\mathcal{B}_{w}$ in complex form is written as:
\begin{subequations}
	\label{wire_em_eq}
	\begin{equation}
		\mathcal{B}_{w}=\mathcal{B}_{wy}+\mathrm{i} \mathcal{B}_{wx}=\mathcal{L}_w f_w \delta_D(s-s_0)	
		\label{b_field},
	\end{equation}
	\begin{equation}
		\mathcal{L}_w=\frac{\mu_0 I_w}{2\pi}
		\label{coef_w},
	\end{equation}
	\begin{equation}
		f_w=\frac{1}{\mathcal{Z}_r+\mathcal{Z}_{wb}}=\sum_{u=0}^{\infty} f_w^{(u)}=\sum_{u=0}^{\infty}\frac{(-1)^u}{\mathcal{Z}_{wb}^{u+1}} \mathcal{Z}_r^u=\sum_{u=0}^{\infty}c_u \mathcal{Z}_r^u,
		\label{fw}
	\end{equation}
\end{subequations}
where $I_w=\mathcal{J}_w L_w$ is the integrated current, $\mathcal{J}_w$ is the wire current, the $\mathcal{Z}_r=x+\mathrm{i} y$ is the test particle position measured from the weak beam and $\mathcal{Z}_{wb}=x_{wb}+\mathrm{i} y_{wb}$ is the position of the weak beam measured from the wire. Since $\mathcal{Z}_{wb}$ must be quite larger that the $\mathcal{Z}_{r}$, the function $f_w$ can be also expressed in a multipolar series as shown in Eq.~(\ref{fw}) with multipole strength $c_u=\frac{(-1)^u}{\mathcal{Z}_{wb}^{u+1}}$. 

\begin{figure}[!thp]
	\subfloat[\label{b_w}]{%
		\includegraphics[width=0.49\textwidth]{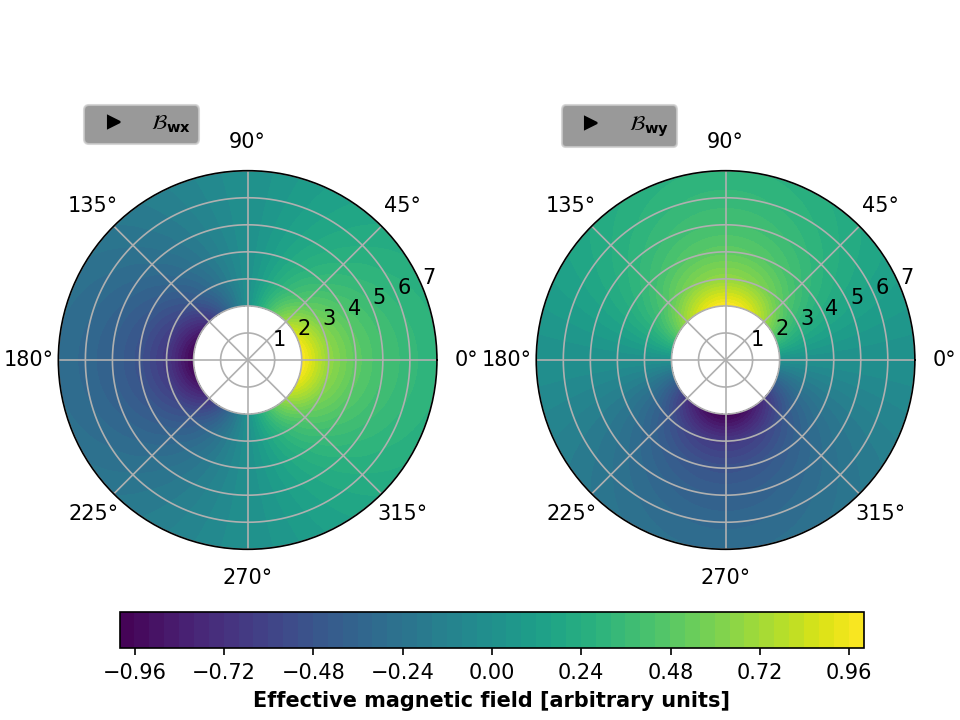}%
	}
	\subfloat[\label{b_bblr}]{%
		\includegraphics[width=0.49\textwidth]{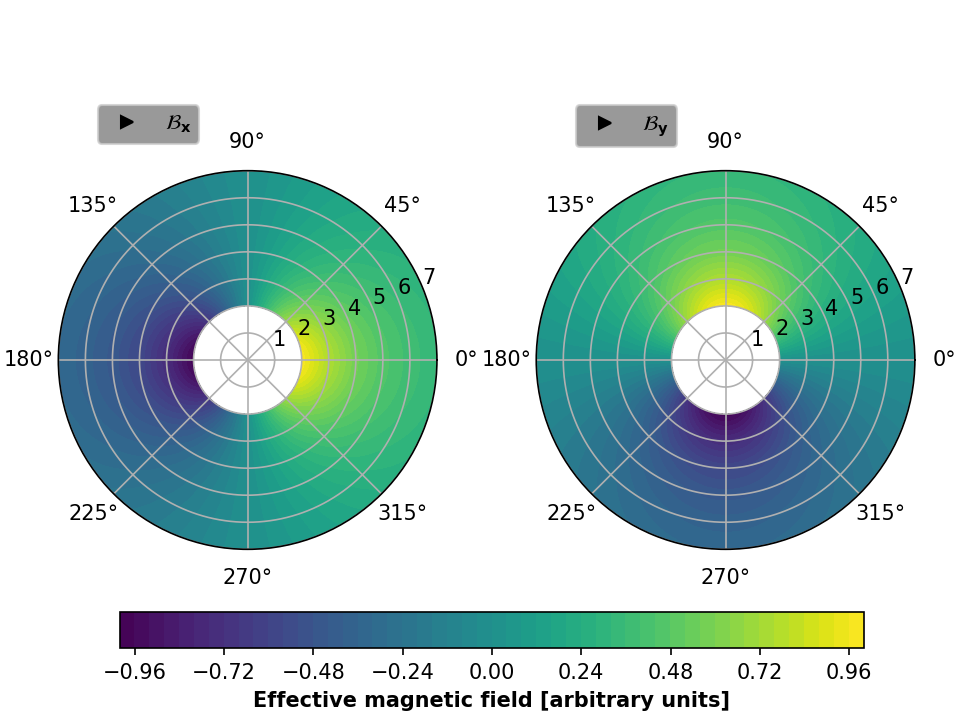}%
	}
	\caption {\label{b_w_bblr}Transverse effective magnetic field from: a) a wire compensator and b) a charge distribution with $\sigma_x>\sigma_y$.}
\end{figure}

The effective magnetic field generated by a bunch in the strong beam ($\mathcal{B}$) and the one generated by a wire compensator ($\mathcal{B}_w$) are quite similar away from their sources. This can be seen in Figs.~\ref{b_w_bblr} where the magnetic fields at distances larger than $2~\sigma$ from their sources $\Big( \sqrt{(\chi+\widetilde{\chi})^2+(\psi+\widetilde{\psi})^2}\ge2~\sigma$ and $\sqrt{\mathtt{Re}\left[\mathcal{Z}_r+\mathcal{Z}_{wb} \right]^2+\mathtt{Im}\left[\mathcal{Z}_r+\mathcal{Z}_{wb} \right]^2 } \ge2~\sigma \Big) $ are plotted. The $\mathcal{B}_{wx}$ and $\mathcal{B}_{wy}$ are shown  in Fiq.~\ref{b_w} while the $\mathcal{B}_x$ and $\mathcal{B}_y$ for a charge distribution with $\sigma_x>\sigma_y$ can be seen in Fig.~\ref{b_bblr}.

Any magnetic field with only transverse components can be derived from a vector potential $A=A_s(x,y)$  using the equations $\vec{B}=\nabla\times \vec{A}$ and $\nabla\cdot \vec{A}=0$. Based on this, the Eqs.~(\ref{bblr_em_eq} and \ref{wire_em_eq}) can be obtained from an effective vector potential $\mathcal{A}_{s}=\mathcal{A}_{s}(x,y)$ according to the formulas $\mathcal{B}_{x}=\frac{d\mathcal{A}_{s}}{dy}$ and $\mathcal{B}_{y}=-\frac{d\mathcal{A}_{s}}{dx}$. Therefore, the Hamiltonian that describes the BBLR and wire kicks is of the form $H=-\frac{q}{P_0}\mathcal{A}_{s}$ and the solution of Hamilton equations are written as:
\begin{subequations}
	\label{bblr_kick}
	\begin{equation}
	\frac{dx}{ds}=0 \Rightarrow x=x^i
	\end{equation}
	\begin{equation}
	\frac{dp_x}{ds}=-\frac{q}{P_0}\mathcal{B}_{y} \Rightarrow p_{x}=p_{x}^i-\frac{q}{P_0}\mathcal{B}_{y}(x^i,y^i;s_0)
	\label{bblr_kick_px}
	\end{equation}
	\begin{equation}
	\frac{dy}{ds}=0 \Rightarrow y=y^i
	\end{equation}
	\begin{equation}
	\frac{dp_y}{ds}=\frac{q}{P_0}\mathcal{B}_{x} \Rightarrow p_{y}=p_{y}^i+\frac{q}{P_0}\mathcal{B}_{x}(x^i,y^i;s_0)
	\label{bblr_kick_py}
	\end{equation}
	\begin{equation}
	\frac{dl}{ds}=0 \Rightarrow l=l^i
	\end{equation}
	\begin{equation}
	\frac{d\delta}{ds}=0 \Rightarrow \delta=\delta^i
	\end{equation}
\end{subequations}
where the effective fields are calculated at $s_0$ and the subscript $i$ denote the initial values. In a circular accelerator like the HL-LHC, the particles of the weak beam experience the BBLR and wire kicks (described in Eqs.~(\ref{bblr_kick})) at each revolution at specific positions. In other words, each of these kicks is periodic with a period equal to the revolution frequency and for that reason different resonances can be excited. The strength of the resonances (RDT$_\chi$) driven by the different types of the BBLR kicks ($\sigma_x>\sigma_y$, $\sigma_y>\sigma_x$ and $\sigma_x=\sigma_y=\sigma$) and from the wire kick are calculated in the following sections.

\section{Resonance driving terms}
Using the Eqs.~(\ref{bblr_em_eq} \ref{wire_em_eq}, \ref{bblr_kick_px} and \ref{bblr_kick_py}) the strength of the resonances (RDT$_\chi$) exited by the BBLR and wire kicks is calculated. In the following calculations the contribution from a non zero $\delta$ and the desertion (coupling of the transverse motion with the longitudinal one) is not explicitly expressed. However, in order to extract this information the action $J_{\chi}$ can be replaced by $\frac{\left( \sqrt{\frac{2\beta_{\chi}J_\chi}{\widetilde{\delta}}}+\frac{D_\chi \delta+\mathcal{O}_{\chi_p}\left( \delta^2\right) }{\mathtt{Cos}\left[ \phi_\chi\right]}\right)^2 }{2\beta_{\chi}}$ with $\chi=x,y$.

\subsection{RDT$_\chi$ driven by elliptical bunches with $\sigma_x>\sigma_y$}
For elliptical strong bunches with $\sigma_x>\sigma_y$, the transverse momentum deviation is defined by the following expressions:
\begin{subequations}
	\begin{equation}
	\frac{dp_x}{ds}=\frac{q}{P_0}\frac{\beta_{ls}}{c}\mathcal{L}_xIm[f_x] \delta_D(s-s_0)	
	\end{equation}
	\begin{equation}
	\frac{dp_y}{ds}=\frac{q}{P_0}\frac{\beta_{ls}}{c}\mathcal{L}_xRe[f_x] \delta_D(s-s_0)			\end{equation}
\end{subequations}
where the $\mathcal{L}_x$ and $f_x$ are equal to $\mathcal{L}$ and $f$ (Eqs.~(\ref{field_f} and \ref{field_L})) when $\sigma_x>\sigma_y$. The function $f_x$ after a Taylor series expansions around the complex numbers $z=x+\mathrm{i} y$, $z_1=y\sigma_x^2-\mathrm{i} x\sigma_y^2$ and $z_2=\mathrm{i} x-y$ is written as:
\begin{equation}
f_x=\sum_{u=0}^{\infty}c_u (x+\mathrm{i} y)^u\left(\sum_{n=0}^{\infty}g_n\left( \frac{y \sigma_x^2 -\mathrm{i} x \sigma_y^2}{\sigma_x\sigma_y\varDelta_x}\right)^n  + \sum_{m=0}^{\infty}w_m\left( \frac{\mathrm{i} x-y}{\varDelta_x}\right)^m \right) 
\end{equation}
Each of the coefficients $c_u$, $g_n$ and $w_m$ form a holonomic sequence and are given by the following recursive relations:
\begin{subequations}
	\label{coef_elxy}
	\begin{equation}
	c_0=\mathtt{Exp}\left[ -\left( \frac{\widetilde{x}+\mathrm{i} \widetilde{y}}{\varDelta_x}\right)^2 \right] 
	\end{equation}
	\begin{equation}
	c_1=-\frac{2 c_0 }{\varDelta_x^2} (\widetilde{x}+\mathrm{i} \widetilde{y}) 
	\end{equation}
	\begin{equation}
	c_u=\frac{-2c_{u-2}-2(\widetilde{x}+\mathrm{i} \widetilde{y})c_{u-1}}{u \varDelta_x^2}~~\text{for}~u=2,3,...
	\end{equation}
	\begin{equation}
	g_0=\mathtt{Erf}\left[ \frac{\widetilde{y} \sigma_x^2-\mathrm{i} \widetilde{x} \sigma_y^2}{\sigma_x\sigma_y\varDelta_x}\right] 
	\end{equation}
	\begin{equation}
	g_1=\frac{2}{\sqrt{\pi}}\mathtt{Exp}\left[ -\left( \frac{\widetilde{y} \sigma_x^2-\mathrm{i} \widetilde{x} \sigma_y^2}{\sigma_x\sigma_y\varDelta_x}\right)^2 \right] 
	\end{equation}
	\begin{equation}
	g_n=\frac{-2\sigma_x\sigma_y\varDelta_x(n-2) g_{n-2}-2(\widetilde{y} \sigma_x^2-\mathrm{i} \widetilde{x} \sigma_y^2)(n-1)g_{n-1}}{n(n-1)\sigma_x\sigma_y\varDelta_x}~~\text{for}~n=2,3,...
	\end{equation}
	\begin{equation}
	w_0=\mathtt{Erf}\left[ \frac{\mathrm{i} \widetilde{x}-\widetilde{y}}{\varDelta_x}\right] 
	\end{equation}
	\begin{equation}
	w_1=\frac{2}{\sqrt{\pi}}\mathtt{Exp}\left[ -\left( \frac{\mathrm{i} \widetilde{x}-\widetilde{y}}{\varDelta_x}\right)^2 \right] 
	\end{equation}
	\begin{equation}
	w_m=\frac{-2\varDelta_x(m-2) w_{m-2}-2(\mathrm{i} \widetilde{x}-\widetilde{y})(m-1)w_{m-1}}{m(m-1)\varDelta_x}~~\text{for}~m=2,3,....
	\end{equation}
\end{subequations}
Moving from the coordinates ($x,y$) to the action angle variables ($J_x,\phi_x,J_y,\phi_y$) according to the transformations $x=\sqrt{2\beta_xJ_x} \mathtt{Cos}\left[ \phi_x\right] $ and $y=\sqrt{2\beta_yJ_y} \mathtt{Cos}\left[ \phi_y\right] $, the $f_x$ is defined by:
\begin{subequations}
	\label{fxunm}
	\begin{equation}
	f_x=\sum_{u=0}^{\infty}\sum_{n=0}^{\infty}\sum_{m=0}^{\infty}f_x^{(u,n,m)}
	\end{equation}
	\begin{equation}
	\begin{split}
	f_x^{(u,n,m)}=&\frac{c_u g_n}{(\sigma_x\sigma_y\varDelta_x)^n} \sum_{l_0=0}^{u}\sum_{l_1=0}^{n}\binom{u}{l_0}\binom{n}{l_1} \left( \sqrt{2\beta_xJ_x}\right)^{l_0} \left( -\mathrm{i} \sigma_y^2 \sqrt{2\beta_xJ_x}\right)^{l_1} \left( \mathrm{i} \sqrt{2\beta_yJ_y}\right)^{u-l_0} ~\times \\ 
	& \left( \sigma_x^2 \sqrt{2\beta_yJ_y} \right)^{n-l_1} \left( \mathtt{Cos}\left[ \phi_x\right]\right)^{l_0+l_1} \left( \mathtt{Cos}\left[ \phi_y\right] \right)^{u+n-(l_0+l_1)} ~+ \\   
	& \frac{c_u w_m}{\varDelta_x^m} \sum_{l_0=0}^{u}\sum_{l_2=0}^{m}\binom{u}{l_0}\binom{m}{l_2} \left( \sqrt{2\beta_xJ_x}\right)^{l_0}  \left( \mathrm{i} \sqrt{2\beta_xJ_x}\right)^{l_2} \left( \mathrm{i} \sqrt{2\beta_yJ_y}\right)^{u-l_0} ~\times \\ 
	&\left( -\sqrt{2\beta_yJ_y} \right)^{m-l_2} \left( \mathtt{Cos}\left[ \phi_x\right]\right)^{l_0+l_2} \left( \mathtt{Cos}\left[ \phi_y\right]\right)^{u+m-(l_0+l_2)}.
	\end{split}
	\end{equation}
\end{subequations}
Since the $f_x$ and so the $f_x^{(u,n,m)}$ are periodic with period of one revolution (the different BBLR kicks are applied at every revolution at specific positions) it is more convenient to use a new set of canonical variables where the new angle $\mu$ will be change linearly with $s$ for the linear motion. Using a generating function of the second type~\cite{gold} the transformation to the new canonical conjugate variables is given by the following equations:
\begin{subequations}
	\label{newjphi}
	\begin{equation}
	\mathcal{F}_2(\phi_x,\widehat{J}_x,\phi_y,\widehat{J}_y,s)=\widehat{J}_x\left(\phi_x + \frac{2\pi Q_x s}{\mathcal{C}} - \int_{0}^{s}\frac{ds'}{\beta_x(s')}\right) +  \widehat{J}_y\left(\phi_y + \frac{2\pi Q_y s}{\mathcal{C}} - \int_{0}^{s}\frac{ds'}{\beta_y(s')}\right)
	\end{equation}
	\begin{equation}
	J_x=\frac{\partial \mathcal{F}_2}{\partial \phi_x}=\widehat{J}_x
	\end{equation}
	\begin{equation}
	\mu_x=\frac{\partial \mathcal{F}_2}{\partial \widehat{J}_x}=\phi_x + \frac{2\pi Q_x s}{\mathcal{C}} - \int_{0}^{s}\frac{ds'}{\beta_x(s')}=\phi_x + \psi_x(s)
	\end{equation}
	\begin{equation}
	J_y=\frac{\partial \mathcal{F}_2}{\partial \phi_y}=\widehat{J}_y
	\end{equation}
	\begin{equation}
	\mu_y=\frac{\partial \mathcal{F}_2}{\partial \widehat{J}_y}=\phi_y + \frac{2\pi Q_y s}{\mathcal{C}} - \int_{0}^{s}\frac{ds'}{\beta_y(s')}=\phi_y + \psi_y(s).
	\end{equation}
\end{subequations}
The $Q_x$ and $Q_y$ are the working tunes, $\beta_x$ and $\beta_y$ are the optical betatronic functions, $\mathcal{C}$ is the lattice circumference and $\psi_\chi(s)=\frac{2\pi Q_\chi s}{\mathcal{C}} - \int_{0}^{s}\frac{ds'}{\beta_\chi(s')}$ with $\chi=x,y$. Using the above transformation and rewriting the cosines in complex form the $f_x^{(u,n,m)}$ is written as:
\begin{subequations}
	\begin{equation}
	\begin{split}
	f_x^{(u,n,m)}=&\sum_{l_0=0}^{u}\sum_{l_1=0}^{n}\sum_{\kappa_1=0}^{\zeta_1}\sum_{\tau_1=0}^{\xi_1} J_x^{\zeta_1/2} ~J_y^{\xi_1/2}~h_{l_0,l_1,\kappa_1,\tau_1}^{(u,n)}  \mathtt{Exp}\left[ \mathrm{i} \left( \mu_x (2\kappa_1-\zeta_1)+\mu_y (2\tau_1-\xi_1)\right) \right] ~+ \\
	&\sum_{l_0=0}^{u}\sum_{l_2=0}^{m}\sum_{\kappa_2=0}^{\zeta_2}\sum_{\tau_2=0}^{\xi_2} J_x^{\zeta_2/2} ~J_y^{\xi_2/2}~h_{l_0,l_2,\kappa_2,\tau_2}^{(u,m)}  \mathtt{Exp}\left[ \mathrm{i}\left( \mu_x (2\kappa_2-\zeta_2)+\mu_y (2\tau_2-\xi_2)\right) \right] 
	\end{split}
	\end{equation}
	\begin{equation}
	\begin{split}
	h_{l_0,l_1,\kappa_1,\tau_1}^{(u,n)}=& \frac{c_u g_n}{(\sigma_x\sigma_y\varDelta_x)^n}\frac{1}{2^{\zeta_1+\xi_1}} \binom{u}{l_0}\binom{n}{l_1}\binom{\zeta_1}{\kappa_1}\binom{\xi_1}{\tau_1} \left( \sqrt{2\beta_x}\right)^{l_0} \left( -\mathrm{i} \sigma_y^2 \sqrt{2\beta_x}\right)^{l_1}  \left( \mathrm{i} \sqrt{2\beta_y}\right)^{u-l_0} ~\times \\
	&\left( \sigma_x^2 \sqrt{2\beta_y} \right)^{n-l_1} \mathtt{Exp}\left[ -\mathrm{i}\left( \psi_x (2\kappa_1-\zeta_1) +\psi_y (2\tau_1-\xi_1)\right) \right]
	\end{split}
	\end{equation}
	\begin{equation}
	\begin{split}
	h_{l_0,l_2,\kappa_2,\tau_2}^{(u,m)}=& \frac{c_u w_m}{\varDelta_x^m}\frac{1}{2^{\zeta_2+\xi_2}} \binom{u}{l_0}\binom{m}{l_2}\binom{\zeta_2}{\kappa_2}\binom{\xi_2}{\tau_2} \left( \sqrt{2\beta_x}\right)^{l_0} \left( \mathrm{i} \sqrt{2\beta_x}\right)^{l_2} \left( \mathrm{i} \sqrt{2\beta_y}\right)^{u-l_0} ~\times \\
	&\left( -\sqrt{2\beta_y} \right)^{m-l_2} \mathtt{Exp}\left[ -\mathrm{i}\left( \psi_x (2\kappa_2-\zeta_2) +\psi_y (2\tau_2-\xi_2)\right) \right].
	\end{split}
	\end{equation}
	\label{fx_expanded}
\end{subequations}
with $\zeta_1=l_0+l_1$, $\xi_1=u+n-\zeta_1$, $\zeta_2=l_0+l_2$, $\xi_2=u+m-\zeta_2$ and the optical parameters are kept together in the $h_{l_0,l_1,\kappa_1,\tau_1}^{(u,n)}$ and $h_{l_0,l_2,\kappa_2,\tau_2}^{(u,m)}$ functions. As said, the BBLR kicks affect the beam at every revolution thus, the $f_x^{(u,n,m)}$ can be expressed as a Fourier series. The resulted series with its Fourier coefficients $h_{l_0,l_1,\kappa_1,\tau_1;p1}^{(u,n)}$ and $h_{l_0,l_2,\kappa_2,\tau_2;p2}^{(u,m)}$ are given by: 
\begin{subequations}
	\begin{equation}
	\begin{split}
	f_x^{(u,n,m)}=&\sum_{l_0=0}^{u}\sum_{l_1=0}^{n}\sum_{\kappa_1=0}^{\zeta_1}\sum_{\tau_1=0}^{\xi_1}\sum_{p_1=-\infty}^{\infty} J_x^{\zeta_1/2} ~J_y^{\xi_1/2}~h_{l_0,l_1,\kappa_1,\tau_1;p_1}^{(u,n)} ~\times \\
	&\mathtt{Exp}\left[ \mathrm{i} \left( \mu_x (2\kappa_1-\zeta_1)+\mu_y (2\tau_1-\xi_1) - \frac{2\pi~s}{\mathcal{C}}p_1 \right) \right] ~+ \\
	&\sum_{l_0=0}^{u}\sum_{l_2=0}^{m}\sum_{\kappa_2=0}^{\zeta_2}\sum_{\tau_2=0}^{\xi_2}\sum_{p_2=-\infty}^{\infty} J_x^{\zeta_2/2} ~J_y^{\xi_2/2}~h_{l_0,l_2,\kappa_2,\tau_2;p_2}^{(u,m)} ~\times \\
	&\mathtt{Exp}\left[ \mathrm{i}\left( \mu_x (2\kappa_2-\zeta_2)+\mu_y (2\tau_2-\xi_2) - \frac{2\pi~s}{\mathcal{C}}p_2 \right) \right]
	\end{split}
	\end{equation}
	\begin{equation}
	h_{l_0,l_1,\kappa_1,\tau_1;p_1}^{(u,n)}= \frac{1}{2\pi}\int_{s_0}^{s_0+\mathcal{C}}  	h_{l_0,l_1,\kappa_1,\tau_1}^{(u,n)} \mathtt{Exp}\left[ \mathrm{i} p_1~\frac{2\pi~s}{\mathcal{C}} \right]~ds
	\end{equation}
	\begin{equation}
	h_{l_0,l_2,\kappa_2,\tau_2;p_2}^{(u,m)}= \frac{1}{2\pi}\int_{s_0}^{s_0+\mathcal{C}}  	h_{l_0,l_2,\kappa_2,\tau_2}^{(u,m)} \mathtt{Exp}\left[ \mathrm{i} p_2~\frac{2\pi~s}{\mathcal{C}} \right]~ds
	\end{equation}
\end{subequations}
The exited resonances are the ones described by the formulas $(2\kappa_1-\zeta_1)Q_x  +(2\tau_1-\xi_1)Q_y =p_1$ and $(2\kappa_2-\zeta_2)Q_x +(2\tau_2-\xi_2)Q_y =p_2$ and their strength at $x$ and $y$ plane are defined by:
\begin{subequations}
	\begin{equation}
	\text{RDT}_x=
	\begin{cases}
	Im\left[ \frac{q}{P_0}\frac{\beta_{ls}}{c}\mathcal{L}_x~h_{l_0,l_1,\kappa_1,\tau_1;p_1}^{(u,n)} \right]  ~~\text{for}\ (2\kappa_1-\zeta_1)Q_x  +(2\tau_1-\xi_1)Q_y =p_1 \\
	Im\left[ \frac{q}{P_0}\frac{\beta_{ls}}{c}\mathcal{L}_x~h_{l_0,l_2,\kappa_2,\tau_2;p_2}^{(u,m)} \right]  ~~\text{for}\ (2\kappa_2-\zeta_2)Q_x +(2\tau_2-\xi_2)Q_y =p_2
	\end{cases}
	\end{equation}
	\begin{equation}
	\text{RDT}_y=
	\begin{cases}
	Re\left[ \frac{q}{P_0}\frac{\beta_{ls}}{c}\mathcal{L}_x~h_{l_0,l_1,\kappa_1,\tau_1;p_1}^{(u,n)} \right] ~~\text{for}\ (2\kappa_1-\zeta_1)Q_x  +(2\tau_1-\xi_1)Q_y =p_1 \\
	Re\left[ \frac{q}{P_0}\frac{\beta_{ls}}{c}\mathcal{L}_x~h_{l_0,l_2,\kappa_2,\tau_2;p_2}^{(u,m)} \right] ~~\text{for}\ (2\kappa_2-\zeta_2)Q_x +(2\tau_2-\xi_2)Q_y =p_2.
	\end{cases}
	\end{equation}
\end{subequations}

\subsection{RDT$_\chi$ driven by elliptical bunches with $\sigma_y>\sigma_x$} 
	
For elliptical strong bunches with $\sigma_y>\sigma_x$ the transverse momentum deviation is defined by the following expressions:
\begin{subequations}
	\begin{equation}
	\frac{dp_x}{ds}=\frac{q}{P_0}\frac{\beta_{ls}}{c}\mathcal{L}_yRe[f_y] \delta(s-s_0)	
	\end{equation}
	\begin{equation}
	\frac{dp_y}{ds}=\frac{q}{P_0}\frac{\beta_{ls}}{c}\mathcal{L}_yIm[f_y] \delta(s-s_0)	
	\end{equation}
\end{subequations}
where the $\mathcal{L}_y$ and $f_y$ are equal to $\mathcal{L}$ and $f$ (Eqs.~(\ref{field_f} and \ref{field_L})) when $\sigma_y>\sigma_x$. The $f_y$ after a Taylor series expansions and the transformation to action angle variable ($\widehat{J}_x,\mu_x,\widehat{J}_y,\mu_y$), is given by the corresponding expressions for the $f_x$ (Eqs.~(\ref{fx_expanded})) provided that the alternation between $x$ and $y$ is performed and is written as:
\begin{subequations}
	\begin{equation}
	\begin{split}
	f_y^{(u,n,m)}=&\sum_{l_0=0}^{u}\sum_{l_1=0}^{n}\sum_{\kappa_1=0}^{\zeta_1}\sum_{\tau_1=0}^{\xi_1} J_y^{\zeta_1/2} ~J_x^{\xi_1/2}~h1_{l_0,l_1,\kappa_1,\tau_1}^{(u,n)}  \mathtt{Exp}\left[ \mathrm{i} \left( \mu_y (2\kappa_1-\zeta_1)+\mu_x (2\tau_1-\xi_1)\right) \right] ~+  \\
	&\sum_{l_0=0}^{u}\sum_{l_2=0}^{m}\sum_{\kappa_2=0}^{\zeta_2}\sum_{\tau_2=0}^{\xi_2} J_y^{\zeta_2/2} ~J_x^{\xi_2/2}~h1_{l_0,l_2,\kappa_2,\tau_2}^{(u,m)}  \mathtt{Exp}\left[ \mathrm{i}\left( \mu_y (2\kappa_2-\zeta_2)+\mu_x (2\tau_2-\xi_2)\right) \right] 
	\end{split}
	\end{equation}
	\begin{equation}
	\begin{split}
	h1_{l_0,l_1,\kappa_1,\tau_1}^{(u,n)}=& \frac{c_u g_n}{(\sigma_x\sigma_y\varDelta_y)^n}\frac{1}{2^{\zeta_1+\xi_1}} \binom{u}{l_0}\binom{n}{l_1}\binom{\zeta_1}{\kappa_1}\binom{\xi_1}{\tau_1} \left( \sqrt{2\beta_y}\right)^{l_0} \left( -\mathrm{i} \sigma_x^2 \sqrt{2\beta_y}\right)^{l_1}  \left( \mathrm{i} \sqrt{2\beta_x}\right)^{u-l_0} ~\times \\
	&\left( \sigma_y^2 \sqrt{2\beta_x} \right)^{n-l_1} \mathtt{Exp}\left[ -\mathrm{i}\left( \psi_y (2\kappa_1-\zeta_1) +\psi_x (2\tau_1-\xi_1)\right) \right]
	\end{split}
	\end{equation}
	\begin{equation}
	\begin{split}
	h1_{l_0,l_2,\kappa_2,\tau_2}^{(u,m)}=& \frac{c_u w_m}{\varDelta_y^m}\frac{1}{2^{\zeta_2+\xi_2}} \binom{u}{l_0}\binom{m}{l_2}\binom{\zeta_2}{\kappa_2}\binom{\xi_2}{\tau_2} \left( \sqrt{2\beta_y}\right)^{l_0} \left( \mathrm{i} \sqrt{2\beta_y}\right)^{l_2} \left( \mathrm{i} \sqrt{2\beta_x}\right)^{u-l_0} ~\times \\
	&\left( -\sqrt{2\beta_x} \right)^{m-l_2} \mathtt{Exp}\left[ -\mathrm{i}\left( \psi_y (2\kappa_2-\zeta_2) +\psi_x (2\tau_2-\xi_2)\right) \right]
	\end{split}
	\end{equation}
	\begin{equation}
	c_0=\mathtt{Exp}\left[ -\left( \frac{\widetilde{y}+\mathrm{i} \widetilde{x}}{\varDelta_y}\right)^2 \right] 
	\label{coef_elyx_first}
	\end{equation}
	\begin{equation}
	c_1=-\frac{2~c_0 }{\varDelta_y^2} (\widetilde{y}+\mathrm{i} \widetilde{x}) 
	\end{equation}
	\begin{equation}
	\text{for}~u=2,3,...~~c_u=\frac{-2c_{u-2}-2(\widetilde{y}+\mathrm{i} \widetilde{x})c_{u-1}}{u~\varDelta_y^2}
	\end{equation}
	\begin{equation}
	g_0=\mathtt{Erf}\left[ \frac{\widetilde{x} \sigma_y^2-\mathrm{i} \widetilde{y} \sigma_x^2}{\sigma_x\sigma_y\varDelta_y}\right] 
	\end{equation}
	\begin{equation}
	g_1=\frac{2}{\sqrt{\pi}}\mathtt{Exp}\left[ -\left( \frac{\widetilde{x} \sigma_y^2-\mathrm{i} \widetilde{y} \sigma_x^2}{\sigma_x\sigma_y\varDelta_y}\right)^2 \right] 
	\end{equation}
	\begin{equation}
	\text{for}~n=2,3,... ~~ g_n=\frac{-2\sigma_x\sigma_y\varDelta_y(n-2) q_{n-2}-2(\widetilde{x} \sigma_y^2-\mathrm{i} \widetilde{y} \sigma_x^2)(n-1)g_{n-1}}{n(n-1)\sigma_x\sigma_y\varDelta_y}
	\end{equation}
	\begin{equation}
	w_0=\mathtt{Erf}\left[ \frac{\mathrm{i} \widetilde{y}-\widetilde{x}}{\varDelta_y}\right] 
	\end{equation}
	\begin{equation}
	w_1=\frac{2}{\sqrt{\pi}}\mathtt{Exp}\left[ -\left( \frac{\mathrm{i} \widetilde{y}-\widetilde{x}}{\varDelta_y}\right)^2 \right] 
	\end{equation}
	\begin{equation}
	\text{for}~m=2,3,... ~~ w_m=\frac{-2\varDelta_y(m-2) w_{m-2}-2(\mathrm{i} \widetilde{y}-\widetilde{x})(m-1)w_{m-1}}{m(m-1)\varDelta_y}.
	\label{coef_elyx_last}
	\end{equation}
\end{subequations}	
The RDT$_\chi$ after the Fourier transformation are defined by: 
\begin{subequations}
	\begin{equation}
	\text{RDT}_x=
	\begin{cases}
	Re\left[ \frac{q}{P_0}\frac{\beta_{ls}}{c}\mathcal{L}_y ~h1_{l_0,l_1,\kappa_1,\tau_1;p_1}^{(u,n)} \right]  ~~\text{for}\ (2\kappa_1-\zeta_1)Q_y  +(2\tau_1-\xi_1)Q_x =p_1 \\
	Re\left[ \frac{q}{P_0}\frac{\beta_{ls}}{c}\mathcal{L}_y ~h1_{l_0,l_2,\kappa_2,\tau_2;p_2}^{(u,m)} \right]  ~~\text{for}\ (2\kappa_2-\zeta_2)Q_y +(2\tau_2-\xi_2)Q_x =p_2
	\end{cases}
	\end{equation}
	\begin{equation}
	\text{RDT}_y=
	\begin{cases}
	Im\left[ \frac{q}{P_0}\frac{\beta_{ls}}{c}\mathcal{L}_y~h1_{l_0,l_1,\kappa_1,\tau_1;p_1}^{(u,n)} \right] ~~\text{for}\ (2\kappa_1-\zeta_1)Q_y  +(2\tau_1-\xi_1)Q_x =p_1 \\
	Im\left[ \frac{q}{P_0}\frac{\beta_{ls}}{c}\mathcal{L}_y~h1_{l_0,l_2,\kappa_2,\tau_2;p_2}^{(u,m)} \right] ~~\text{for}\ (2\kappa_2-\zeta_2)Q_y +(2\tau_2-\xi_2)Q_x =p_2
	\end{cases}
	\end{equation}
	\begin{equation}
	h1_{l_0,l_1,\kappa_1,\tau_1;p_1}^{(u,n)}= \frac{1}{2\pi}\int_{s_0}^{s_0+\mathcal{C}}  	h1_{l_0,l_1,\kappa_1,\tau_1}^{(u,n)} \mathtt{Exp}\left[ \mathrm{i} p_1~\frac{2\pi~s}{\mathcal{C}} \right]~ds
	\end{equation}
	\begin{equation}
	h1_{l_0,l_2,\kappa_2,\tau_2;p_2}^{(u,m)}= \frac{1}{2\pi}\int_{s_0}^{s_0+\mathcal{C}}  	h1_{l_0,l_2,\kappa_2,\tau_2}^{(u,m)} \mathtt{Exp}\left[ \mathrm{i} p_2~\frac{2\pi~s}{\mathcal{C}} \right]~ds .
	\end{equation}
\end{subequations}

\subsection{RDT$_\chi$ driven by round bunches with $\sigma_x=\sigma_y$}

For round strong bunches ($\sigma_x=\sigma_y=\sigma$) the transverse momentum deflection from a BBLR kick is defined by:
\begin{subequations}
	\begin{equation}
	\frac{dp_x}{ds}=\frac{q}{P_0}\frac{\beta_{ls}}{c}\mathcal{L}_r f_r (x+\widetilde{x}) \delta(s-s_0)
	\end{equation}
	\begin{equation}
	\frac{dp_y}{ds}=\frac{q}{P_0}\frac{\beta_{ls}}{c}\mathcal{L}_r f_r (y+\widetilde{y}) \delta(s-s_0)
	\end{equation}
	\begin{equation}
	f_r=\frac{1}{R^2}\left( 1-\mathtt{Exp}\left[ -\frac{R^2}{2\sigma^2}\right] \right)
	\end{equation}
\end{subequations}
where the $\mathcal{L}_r$ is equal to $\mathcal{L}$ (Eq.~(\ref{field_L})) when $\sigma_x=\sigma_y$ and $R^2=(x+\widetilde{x})^2+(y+\widetilde{y})^2$. After a Taylor series expansions the $f_r$ is given by the following summation:
\begin{equation}
f_r=\sum_{u=0}^{\infty}~f_r^{(u)}=\sum_{u=0}^{\infty}\frac{(-1)^u R^{2u}}{2^{u+1}~(u+1)!~\sigma^{2(u+1)}}= \sum_{u=0}^{\infty}\frac{(-1)^u~\left( (x+\widetilde{x})^2+(y+\widetilde{y})^2\right) ^{u}}{2^{u+1}~(u+1)!~\sigma^{2(u+1)}}.
\end{equation}
Using the action angle variables $\chi=\sqrt{2\beta_\chi J_\chi}~\mathtt{Cos}\left[ \phi_\chi\right] $ with $\chi=x,y$ and expanding the binomials, the $f_r$ takes the following form:
\begin{subequations}
	\label{fr}
	\begin{equation}
	f_r=\sum_{u=0}^{\infty}~f_r^{(u)}
	\end{equation}
	\begin{equation}
	\begin{split}
	f_r^{(u)}=&\sum_{l=0}^{u}\sum_{n=0}^{2l}\sum_{m=0}^{2(u-l)}\frac{(-1)^u}{2^{u+1}~(u+1)!~\sigma^{2(u+1)}}\binom{u}{l}\binom{2l}{n}\binom{2(u-l)}{m} ~\times \\
	&(\sqrt{2\beta_x})^n ~\widetilde{x}^{2l-n} ~(\sqrt{2\beta_y})^m ~\widetilde{y}^{2(u-l)-m} ~\left( \mathtt{Cos}\left[ \phi_x\right]\right) ^n ~\left( \mathtt{Cos}\left[ \phi_y\right]\right) ^m.
	\end{split}
	\end{equation}
\end{subequations}
Rewriting the cosines in complex form ($\mathtt{Cos}\left[ \phi_\chi\right] =\frac{\mathtt{Exp}[\mathrm{i} \phi_\chi]+\mathtt{Exp}[-\mathrm{i} \phi_\chi]}{2}$ with $\chi=x,y$) and using the action angle variables ($\widehat{J}_x,\mu_x,\widehat{J}_y,\mu_y$) described by the Eqs.~(\ref{newjphi}), the $f_{rx}$ and $f_{ry}$ for a given $u$ are written as:
\begin{subequations}
	\begin{equation}
	\begin{split}
	f_{rx}^{(u)}=& (x+\widetilde{x})\sum_{l=0}^{u}f_r^{(u)}= \sum_{l=0}^{u}\sum_{n=0}^{2l}\sum_{m=0}^{2(u-l)}\sum_{\tau=0}^{m}J_x^{n/2}J_y^{m/2}~\mathtt{Exp}\left[ \mathrm{i}  \mu_y (2\tau-m)\right] \times \\
	&\Bigg( \sqrt{J_x}~\sum_{\kappa=0}^{n+1}~h2_{l,n,m,\kappa,\tau}^{(u)} \mathtt{Exp}\left[ \mathrm{i}  \mu_x (2\kappa-n-1) \right]  +  \sum_{\kappa=0}^{n}~h3_{l,n,m,\kappa,\tau}^{(u)} \mathtt{Exp}\left[ \mathrm{i}  \mu_x (2\kappa-n) \right]  ~ \Bigg) 
	\end{split}
	\end{equation}
	\begin{equation}
	\begin{split}
	f_{ry}^{(u)}=&(y+\widetilde{y})\sum_{l=0}^{u}f_r^{(u)}= \sum_{l=0}^{u}\sum_{n=0}^{2l}\sum_{m=0}^{2(u-l)}\sum_{\kappa=0}^{n}J_x^{n/2}J_y^{m/2}~\mathtt{Exp}\left[ \mathrm{i}  \mu_x (2\kappa-n)\right] \times \\
	&\Bigg( \sqrt{J_y}~\sum_{\tau=0}^{m+1}~h4_{l,n,m,\kappa,\tau}^{(u)} \mathtt{Exp}\left[ \mathrm{i}  \mu_y (2\tau-m-1) \right]  + \sum_{\tau=0}^{m}~h5_{l,n,m,\kappa,\tau}^{(u)} \mathtt{Exp}\left[ \mathrm{i}  \mu_y (2\tau-m) \right]  ~ \Bigg) 
	\end{split}
	\end{equation}
	\begin{equation}
	\begin{split}
	h2_{l,n,m,\kappa,\tau}^{(u)}=&\frac{(-1)^u}{2^{u+n+m+2}~(u+1)!~\sigma^{2(u+1)}}\binom{u}{l}\binom{2l}{n}\binom{2(u-l)}{m}\binom{n+1}{\kappa}\binom{m}{\tau} \left( \sqrt{2\beta_x}\right) ^{n+1} ~\widetilde{x}^{2l-n} ~\times \\
	&\left( \sqrt{2\beta_y}\right) ^m ~\widetilde{y}^{2(u-l)-m}  \mathtt{Exp}\left[ -\mathrm{i} \psi_x (2\kappa-n-1) \right] \mathtt{Exp}\left[ -\mathrm{i}  \psi_y (2\tau-m)\right]
	\end{split}
	\end{equation}
	\begin{equation}
	\begin{split}
	h3_{l,n,m,\kappa,\tau}^{(u)}=&\frac{(-1)^u}{2^{u+n+m+1}~(u+1)!~\sigma^{2(u+1)}}\binom{u}{l}\binom{2l}{n}\binom{2(u-l)}{m}\binom{n}{\kappa}\binom{m}{\tau} \left( \sqrt{2\beta_x}\right) ^{n} ~\widetilde{x}^{2l-n+1} ~\times \\
	&\left( \sqrt{2\beta_y}\right) ^m ~\widetilde{y}^{2(u-l)-m} \mathtt{Exp}\left[ -\mathrm{i} \psi_x (2\kappa-n) \right] \mathtt{Exp}\left[ -\mathrm{i}  \psi_y (2\tau-m)\right]
	\end{split}
	\end{equation}
	\begin{equation}
	\begin{split}
	h4_{l,n,m,\kappa,\tau}^{(u)}=&\frac{(-1)^u}{2^{u+n+m+2}~(u+1)!~\sigma^{2(u+1)}}\binom{u}{l}\binom{2l}{n}\binom{2(u-l)}{m}\binom{n}{\kappa}\binom{m+1}{\tau} \left( \sqrt{2\beta_x}\right) ^{n} ~\widetilde{x}^{2l-n} ~\times \\
	&\left( \sqrt{2\beta_y}\right) ^{m+1} ~\widetilde{y}^{2(u-l)-m}  \mathtt{Exp}\left[ -\mathrm{i} \psi_x (2\kappa-n) \right] \mathtt{Exp}\left[ -\mathrm{i}  \psi_y (2\tau-m-1)\right]
	\end{split}
	\end{equation}
	\begin{equation}
	\begin{split}
	h5_{l,n,m,\kappa,\tau}^{(u)}=&\frac{(-1)^u}{2^{u+n+m+1}~(u+1)!~\sigma^{2(u+1)}}\binom{u}{l}\binom{2l}{n}\binom{2(u-l)}{m}\binom{n}{\kappa}\binom{m}{\tau} \left( \sqrt{2\beta_x}\right) ^{n} ~\widetilde{x}^{2l-n} ~\times \\
	&\left( \sqrt{2\beta_y}\right) ^m ~\widetilde{y}^{2(u-l)-m+1}  \mathtt{Exp}\left[ -\mathrm{i} \psi_x (2\kappa-n) \right] \mathtt{Exp}\left[ -\mathrm{i}  \psi_y (2\tau-m)\right].
	\end{split}
	\end{equation}
\end{subequations}
Because of the periodic repetition of the BBLR kicks, the $f_{rx}^{(u)}$ and $f_{ry}^{(u)}$ can be written as a Fourier series. This series with their Fourier coefficients $h2_{l,n,m,\kappa,\tau;p_1}^{(u)}$, $h3_{l,n,m,\kappa,\tau:p_2}^{(u)}$, $h4_{l,n,m,\kappa,\tau;p_1}^{(u)}$ and $h5_{l,n,m,\kappa,\tau;p_2}^{(u)}$ are given by:
\begin{subequations}
	\begin{equation}
	\begin{split}
	f_{rx}^{(u)}=& \sum_{l=0}^{u}\sum_{n=0}^{2l}\sum_{m=0}^{2(u-l)}\sum_{\tau=0}^{m}J_x^{n/2}J_y^{m/2}~\times \\
	&\Bigg( \sqrt{J_x}~\sum_{\kappa=0}^{n+1}\sum_{p_2=-\infty}^{\infty}~h2_{l,n,m,\kappa,\tau;p_1}^{(u)}\mathtt{Exp}\left[ \mathrm{i}  \left( \mu_x (2\kappa-n-1)+\mu_y (2\tau-m) - \frac{2\pi~s}{\mathcal{C}}p_1 \right) \right]  ~+ \\
	& \sum_{\kappa=0}^{n}\sum_{p_1=-\infty}^{\infty}~h3_{l,n,m,\kappa,\tau;p_2}^{(u)} \mathtt{Exp}\left[ \mathrm{i}  \left( \mu_x (2\kappa-n)+\mu_y (2\tau-m) - \frac{2\pi~s}{\mathcal{C}}p_2 \right) \right]  ~ \Bigg) 
	\end{split}
	\end{equation}
	\begin{equation}
	\begin{split}
	f_{ry}^{(u)}=& \sum_{l=0}^{u}\sum_{n=0}^{2l}\sum_{m=0}^{2(u-l)}\sum_{\kappa=0}^{n}J_x^{n/2}J_y^{m/2}~\times \\
	&\Bigg( \sqrt{J_y}~\sum_{\tau=0}^{m+1}\sum_{p_2=-\infty}^{\infty}~h4_{l,n,m,\kappa,\tau;p_1}^{(u)}\mathtt{Exp}\left[ \mathrm{i}  \left( \mu_x (2\kappa-n)+\mu_y (2\tau-m-1) - \frac{2\pi~s}{\mathcal{C}}p_1 \right) \right]  ~+ \\
	& \sum_{\tau=0}^{m}\sum_{p_1=-\infty}^{\infty}~h5_{l,n,m,\kappa,\tau;p_2}^{(u)} \mathtt{Exp}\left[ \mathrm{i}  \left( \mu_x (2\kappa-n)+\mu_y (2\tau-m) - \frac{2\pi~s}{\mathcal{C}}p_2 \right) \right]  ~ \Bigg) 
	\end{split}
	\end{equation}
	\begin{equation}
	h2_{l,n,m,\kappa,\tau;p_1}^{(u)}= \frac{1}{2\pi}\int_{s_0}^{s_0+\mathcal{C}}  	h2_{l,n,m,\kappa,\tau}^{(u)} \mathtt{Exp}\left[ \mathrm{i} p_1~\frac{2\pi~s}{\mathcal{C}} \right]~ds
	\end{equation}
	\begin{equation}
	h3_{l,n,m,\kappa,\tau;p_2}^{(u)}= \frac{1}{2\pi}\int_{s_0}^{s_0+\mathcal{C}}  	h3_{l,n,m,\kappa,\tau}^{(u)} \mathtt{Exp}\left[ \mathrm{i} p_2~\frac{2\pi~s}{\mathcal{C}} \right]~ds
	\end{equation}
	\begin{equation}
	h4_{l,n,m,\kappa,\tau;p_1}^{(u)}= \frac{1}{2\pi}\int_{s_0}^{s_0+\mathcal{C}}  	h4_{l,n,m,\kappa,\tau}^{(u)} \mathtt{Exp}\left[ \mathrm{i} p_1~\frac{2\pi~s}{\mathcal{C}} \right]~ds.
	\end{equation}
	\begin{equation}
	h5_{l,n,m,\kappa,\tau;p_2}^{(u)}= \frac{1}{2\pi}\int_{s_0}^{s_0+\mathcal{C}}  	h5_{l,n,m,\kappa,\tau}^{(u)} \mathtt{Exp}\left[ \mathrm{i} p_2~\frac{2\pi~s}{\mathcal{C}} \right]~ds.
	\end{equation}
\end{subequations}
The exited resonances resulted from $f_{rx}^{(u)}$ are the $Q_x (2\kappa-n-1)+Q_y (2\tau-m) = p_1$ and $Q_x (2\kappa-n)+Q_y (2\tau-m) = p_2$ and from $f_{ry}^{(u)}$ are the $Q_x (2\kappa-n)+Q_y (2\tau-m-1) = p_1$ and $Q_x (2\kappa-n)+Q_y (2\tau-m) = p_2$. The RDT$_\chi$ (resonance strength) at $x$ and $y$ plane is defined by:
\begin{subequations}
	\begin{equation}
	\text{RDT}_x=
	\begin{cases}
	\frac{q}{P_0}\frac{\beta_{ls}}{c}\mathcal{L}_r~h2_{l,n,m,\kappa,\tau;p_1}^{(u)} ~~\text{for}\ Q_x (2\kappa-n-1)+Q_y (2\tau-m) = p_1 \\
	\frac{q}{P_0}\frac{\beta_{ls}}{c}\mathcal{L}_r~h3_{l,n,m,\kappa,\tau;p_2}^{(u)} ~~\text{for}\ Q_x (2\kappa-n)+Q_y (2\tau-m) = p_2
	\end{cases}
	\end{equation}
	\begin{equation}
	\text{RDT}_y=
	\begin{cases}
	\frac{q}{P_0}\frac{\beta_{ls}}{c}\mathcal{L}_r~h4_{l,n,m,\kappa,\tau;p_1}^{(u)} ~~\text{for}\ Q_x (2\kappa-n)+Q_y (2\tau-m-1) = p_1 \\
	\frac{q}{P_0}\frac{\beta_{ls}}{c}\mathcal{L}_r~h5_{l,n,m,\kappa,\tau;p_2}^{(u)} ~~\text{for}\ Q_x (2\kappa-n)+Q_y (2\tau-m) = p_2
	\end{cases}
	\end{equation}
\end{subequations}

\subsection{RDT$_\chi$ driven by a DC wire}
The wire kick change the transverse momentum according to the equations:
\begin{subequations}
	\begin{equation}
	\frac{dp_x}{ds}=-\frac{q}{P_0}\mathcal{L}_w Im\left[ f_{w}\right]=-\frac{q}{P_0}\mathcal{L}_w Im\left[ \sum_{u=0}^{\infty}c_u \mathcal{Z}_r^u \right] \delta(s-s_0)
	\end{equation}
	\begin{equation}
	\frac{dp_y}{ds}=\frac{q}{P_0}\mathcal{L}_w Re\left[ f_{w}\right]  \delta(s-s_0)=\frac{q}{P_0}\mathcal{L}_w Re\left[ \sum_{u=0}^{\infty}c_u \mathcal{Z}_r^u\right]  \delta(s-s_0),
	\end{equation}
\end{subequations}
where the $\mathcal{L}_w$ and $f_w$ are given by the Eqs.~(\ref{coef_w} and \ref{fw}). The complex $\mathcal{Z}_r$ is equal to $\mathcal{Z}_r=x+\mathrm{i} y$ and using the action angle variables $\chi=\sqrt{2\beta_\chi J_\chi} \mathtt{Cos}\left[ \phi_\chi\right] $ with $\chi=x,y$, the $f_w$ for a given $u$ can be written as: 
\begin{equation}
\label{fwu}
\begin{split}
f_w^{(u)}=c_u\sum_{l=0}^{u}\binom{u}{l} (x)^l (\mathrm{i} y)^{u-l}=& c_u\sum_{l=0}^{u}\binom{u}{l}\left( \sqrt{2\beta_xJ_x} \right)^l \left( \mathrm{i}  \sqrt{2\beta_yJ_y}\right)^{u-l} \times \\
&\left( \mathtt{Cos}\left[ \phi_x\right]\right) ^l ~\left( \mathtt{Cos}\left[ \phi_y\right]\right) ^{u-l} .
\end{split}
\end{equation}
Once again using the complex form of the cosine ($\mathtt{Cos}\left[ \phi_\chi\right] =\frac{\mathtt{Exp}[\mathrm{i} \phi_\chi]+\mathtt{Exp}[-\mathrm{i} \phi_\chi]}{2}$ with $\chi=x,y$) and the action angle variables ($\widehat{J}_x,\mu_x,\widehat{J}_y,\mu_y$) according to the Eqs.~(\ref{newjphi}), the $f_{w}^{(u)}$ is given by:
\begin{subequations}
	\begin{equation}
	f_w^{(u)}=\sum_{l=0}^{u}\sum_{\kappa=0}^{l}\sum_{\tau}^{u-l} ~h6^{(u)}_{l,\kappa,\tau} ~J_x^{l/2} ~J_y^{(u-l)/2}  \mathtt{Exp}\left[ \mathrm{i}\left( \mu_x(2\kappa-l)+\mu_y(2\tau+l-u)\right) \right] 
	\end{equation}
	\begin{equation}
	h6^{(u)}_{l,\kappa,\tau}=\frac{c_u}{2^u}~\binom{u}{l}\binom{l}{k}\binom{u-l}{\tau}\left(  \sqrt{2\beta_x}\right)^l\left(\mathrm{i} \sqrt{2\beta_y}\right)^{u-l} \mathtt{Exp}\left[ -\mathrm{i}\left( \psi_x(2\kappa-l)+\psi_y(2\tau+l-u)\right) \right],
	\end{equation}
\end{subequations}	
where all the optical parameters are located in $h6_{l,\kappa,\tau}^{(u)}$. The DC wire kick is experienced by the beam at every revolution thus, the $f_w^{(u)}$ can be expanded in a Fourier series. This series with its Fourier coefficients $h6_{l,\kappa,\tau;p}^{(u,n)}$ are described by: 
\begin{subequations}
	\begin{equation}
	f_w^{(u)}=\sum_{l=0}^{u}\sum_{\kappa=0}^{l}\sum_{\tau}^{u-l}\sum_{p=-\infty}^{\infty} ~h6^{(u)}_{l,\kappa,\tau} ~J_x^{l/2} ~J_y^{(u-l)/2}   \mathtt{Exp}\left[ \mathrm{i}\left( \mu_x(2\kappa-l)+\mu_y(2\tau+l-u)\right) -\frac{2\pi~s}{\mathcal{C}}p\right] 
	\end{equation}
	\begin{equation}
	h6^{(u)}_{l,\kappa,\tau;p}=\frac{1}{2\pi}\int_{s_0}^{s_0+\mathcal{C}}h6^{(u)}_{l,\kappa,\tau} ~\mathtt{Exp}\left[ \mathrm{i} p~ \frac{2\pi~s}{\mathcal{C}}~ds \right].
	\end{equation}
\end{subequations}
\begin{subequations}
	\begin{equation}
	f_w^{(u)}=\sum_{l=0}^{u}\sum_{\kappa=0}^{l}\sum_{\tau}^{u-l}\sum_{p=-\infty}^{\infty} ~h6^{(u)}_{l,\kappa,\tau} ~J_x^{l/2} ~J_y^{(u-l)/2}   \mathtt{Exp}\left[ \mathrm{i}\left( \mu_x(2\kappa-l)+\mu_y(2\tau+l-u)\right) -\frac{2\pi~s}{\mathcal{C}}p\right] 
	\end{equation}
	\begin{equation}
	h6^{(u)}_{l,\kappa,\tau;p}=\frac{1}{2\pi}\int_{s_0}^{s_0+\mathcal{C}}h6^{(u)}_{l,\kappa,\tau} ~\mathtt{Exp}\left[ \mathrm{i} p~ \frac{2\pi~s}{\mathcal{C}}~ds \right].
	\end{equation}
\end{subequations}
The exited resonances are the ones described by the formula $(2\kappa-l)Q_x  +(2\tau+l-u)Q_y =p$ and their strength at $x$ and $y$ plane are defined by:
\begin{subequations}
	\begin{equation}
	\text{RDT}_x=Im\left[ \frac{q}{P_0}\mathcal{L}_w h6^{(u)}_{l,\kappa,\tau;p} \right]
	\end{equation}
	\begin{equation}
	\text{RDT}_y=Re\left[ \frac{q}{P_0}\mathcal{L}_w h6^{(u)}_{l,\kappa,\tau;p} \right].
	\end{equation}
\end{subequations}

\section{Tune spread with amplitude}
Once again, in the following calculations the contribution from a non zero $\delta$ and the desertion (coupling of the transverse motion with the longitudinal one) is not explicitly expressed. However, in order to extract this information the action $J_{\chi}$ can be replaced by $\frac{\left( \sqrt{\frac{2\beta_{\chi}J_\chi}{\widetilde{\delta}}}+\frac{D_\chi \delta+\mathcal{O}_{\chi_p}\left( \delta^2\right) }{\mathtt{Cos}\left[ \phi_\chi\right]}\right)^2 }{2\beta_{\chi}}$ with $\chi=x,y$.

For the calculation of the tine spread with amplitude (TSA$_\chi$), the first perturbation theory is used~\cite{gold}. The particles motion can be described by the following Hamiltonian expressed in action angle variables:
\begin{equation}
H(J,\phi;s)=H_0(J;s)+V(J,\phi;s).
\end{equation}
For our conservative system, the actions $J$ (at the different directions of motion) are calculated using the separable $H_0$. Thus, the actions are constant ($J=const$) and the $H_0$ is independent from the angle variables. The parameter $V$ can be seen as the perturbation that depends on $(J,\phi;s)$ and at the limit where $V\rightarrow0$ it is $H=H_0$. From the Hamilton equations, the tune divination ($\Delta Q_\chi$) due to the perturbation can be calculated according to the equation:
\begin{equation}
\begin{split}
Q_\chi=&\frac{1}{2\pi}\oint d\phi_\chi=\frac{1}{2\pi}\oint \frac{dH}{dJ_\chi}ds \Rightarrow \\
\Delta Q_\chi=&Q_\chi-\frac{1}{2\pi}\oint \frac{dH_0}{dJ_\chi}ds=\frac{1}{2\pi}\oint \frac{dV}{dJ_\chi}ds,
\end{split}
\end{equation}  
where $\chi=x,y$ and the integral is over one revolution.  Keeping only the contribution from the first order energy correction (caused by the perturbation term), the Hamiltonian can be take the simpler form $H=H_0+\left\langle V \right\rangle_\phi $. Using this simplified Hamiltonian, the TSA$_\chi$ is given by:
\begin{equation}
\text{TSA}_\chi= \frac{1}{2\pi}\oint \left\langle\frac{dV}{dJ_\chi}\right\rangle _\phi ds,
\label{tsa}
\end{equation}
where $\left\langle \dots \right\rangle_\phi $ indicates the average over the angles. For $V=-\frac{q}{P_0}\mathcal{A}_{s}$ and using the formulas $\mathcal{B}_{x}=\frac{d\mathcal{A}_{s}}{dy}$ and $\mathcal{B}_{y}=-\frac{d\mathcal{A}_{s}}{dx}$, the derivative of the perturbation (BBLR or wire interaction) over the actions can be written as:
\begin{subequations}
	\label{perd_deriv}
	\begin{equation}
	\frac{dV}{dJ_x}=\frac{dV}{dx}\frac{dx}{dJ_x}=\frac{q}{P_0}\mathcal{B}_{y}\frac{\partial x}{\partial J_x}
	\end{equation}
	\begin{equation}
	\frac{dV}{dJ_y}=\frac{dV}{dy}\frac{dy}{dJ_y}=-\frac{q}{P_0}\mathcal{B}_{x}\frac{\partial y}{\partial J_y}.
	\end{equation}
\end{subequations}
Using the equation $\chi=\sqrt{2\beta_\chi J_\chi} \mathtt{Cos}\left[ \phi_\chi\right] $ with $\chi=x,y$ and the Eqs.~(\ref{tsa} and \ref{perd_deriv}), the TSA$_\chi$ at the two planes is given by:
\begin{subequations}
	\label{tsa_bef}
	\begin{equation}
	\text{TSA}_x= \frac{1}{2\pi}\sqrt{\frac{\beta_x}{2J_x}} \frac{q}{P_0} \left\langle \mathcal{B}_{y}(x,y;s_0)~\mathtt{Cos}\left[ \phi_x\right] \right\rangle_\phi
	\end{equation}
	\begin{equation}
	\text{TSA}_y= -\frac{1}{2\pi}\sqrt{\frac{\beta_y}{2J_y}} \frac{q}{P_0} \left\langle \mathcal{B}_{x}(x,y;s_0)~\mathtt{Cos}\left[ \phi_y\right] \right\rangle_\phi.
	\end{equation}
\end{subequations}		 
For the calculation of the integrals over the angles, the trigonometric redaction formulas are used. For example the trigonometric redaction of a cosine at even ($ev$) or odd ($od$) power is given by:  
\begin{subequations}
	\label{trig_red}
	\begin{equation}
	\mathtt{Cos}^{ev}\left[ \theta\right] =\left\langle \mathtt{Cos}^{ev}\left[ \theta\right]  \right\rangle_\theta+\frac{1}{2^{ev-1}}\sum_{k=0}^{\frac{ev}{2}-1}\binom{ev}{k}\mathtt{Cos}\left[ (ev-2k)\theta\right] 
	\end{equation}
	\begin{equation}
	\left\langle \mathtt{Cos}^{ev}\left[ \theta\right]  \right\rangle_\theta=\frac{1}{2\pi}\int_{0}^{2\pi}\mathtt{Cos}^{ev}\left[ \theta\right] ~d\theta=\frac{1}{2^{ev}}\binom{ev}{\frac{ev}{2}}
	\end{equation}
	\begin{equation}
	\mathtt{Cos}^{od}\left[ \theta\right] =\frac{1}{4^\frac{od-1}{2}}\sum_{k=0}^{\frac{od-1}{2}}\binom{od}{k}\mathtt{Cos}\left[ (od-2k)\theta\right] 
	\end{equation}
	\begin{equation}
	\left\langle \mathtt{Cos}^{od}\left[ \theta\right]  \right\rangle_\theta=\frac{1}{2\pi}\int_{0}^{2\pi}\mathtt{Cos}^{od}\left[ \theta\right] ~d\theta=0.
	\end{equation}
\end{subequations}

\subsection{TSA$_\chi$ driven by elliptical bunches with $\sigma_x>\sigma_y$}
For an elliptical strong bunch with $\sigma_x>\sigma_y$ the Eq.~(\ref{tsa_bef}) take the following form: 
\begin{subequations}
	\label{tsa_fx}
	\begin{equation}
	\text{TSA}_x= -\frac{q~\beta_{ls}}{2\pi ~c ~P_0}\sqrt{\frac{\beta_x}{2J_x}} \mathcal{L}_x Im\left[ \left\langle f_x \mathtt{Cos}\left[ \phi_x\right] \right\rangle_\phi\right] 
	\end{equation}
	\begin{equation}
	\text{TSA}_y= -\frac{q~\beta_{ls}}{2\pi ~c ~P_0}\sqrt{\frac{\beta_y}{2J_y}} \mathcal{L}_x Re\left[ \left\langle f_x \mathtt{Cos}\left[ \phi_y\right] \right\rangle_\phi\right].
	\end{equation}
\end{subequations}	
The $\mathcal{L}_x$ is equal to $\mathcal{L}$ (Eq.~(\ref{field_L})) when $\sigma_x>\sigma_y$ and the $f_x$ is described by the Eqs.~(\ref{coef_elxy} and \ref{fxunm}) calculated at $s_0$. After the integrating over the angles, the TSA$_\chi$ is zero if $u,n$ and $m$ are all even or odd. The two combinations that contribute to the tune spread with amplitude are the ones with [$u$:even, $n$:odd, $m$:odd] and [$u$:odd, $n$:even, $m$:even]. For these cases the TSA$_\chi$ is given by the following equations:
{\footnotesize
\begin{subequations}
	\label{tsa_fxevod}
	\begin{equation}
	\text{TSA}_x= ~\mathrm{i} \frac{q~\beta_{ls}}{2\pi ~c ~P_0} \mathcal{L}_x \Bigg[\sum_{u=0,2,4,\dots}^{\infty}\sum_{n=1,3,5,\dots}^{\infty}\sum_{m=1,3,5,\dots}^{\infty}\mathcal{T}_x^{(u,n,m)} +   \sum_{u=1,3,5,\dots}^{\infty}\sum_{n=0,2,4,\dots}^{\infty}\sum_{m=0,2,4,\dots}^{\infty}\mathcal{T}1_x^{(u,n,m)} \Bigg]
	\end{equation}
	\begin{equation}
	\text{TSA}_y= -\frac{q~\beta_{ls}}{2\pi ~c ~P_0} \mathcal{L}_x \Bigg[\sum_{u=0,2,4,\dots}^{\infty}\sum_{n=1,3,5,\dots}^{\infty}\sum_{m=1,3,5,\dots}^{\infty}\mathcal{T}_y^{(u,n,m)}+  \sum_{u=1,3,5,\dots}^{\infty}\sum_{n=0,2,4,\dots}^{\infty}\sum_{m=0,2,4,\dots}^{\infty}\mathcal{T}1_y^{(u,n,m)} \Bigg]
	\end{equation}
	\begin{equation}
	\begin{split}
	\label{tsa_fxevod_x}
	&\mathcal{T}_x^{(u,n,m)}= \sqrt{\frac{\beta_x}{2J_x}}\Biggg[ \sum_{l=0}^{\frac{u}{2}}\binom{u}{2l}(2\beta_x J_x)^{l}  \left( \mathrm{i}\sqrt{2\beta_y J_y} \right)^{u-2l} \Bigg( \frac{\widetilde{c_u g_n}}{2^{u+n}(\sigma_x\sigma_y\Delta_x)^n} \binom{u-2l}{\frac{u-2l}{2}}\binom{n+2l}{\frac{n+2l-1}{2}} ~\times \\
	&\left( -\mathrm{i}\sqrt{2\beta_x J_x}\sigma_y^2 \right)^n + \frac{\widetilde{c_u w_m}\left( \mathrm{i}\sqrt{2\beta_x J_x} \right)^m}{2^{u+m}\Delta_x^m}\binom{u-2l}{\frac{u-2l}{2}} \binom{m+2l}{\frac{m+2l-1}{2}} -\frac{\mathrm{i}  \widetilde{c_u g_n}\sqrt{2\beta_x J_x}\sigma_y^2 }{2^{u+n}(\sigma_x\sigma_y\Delta_x)^n}  \Bigg( \sum_{k=1}^{\frac{n-3}{2}}\frac{n}{2k+1}\binom{n-1}{2k} ~\times \\
	&\binom{2l+2k+1}{l+k} \binom{u+n-2l-2k-1}{\frac{u+n-2l-2k-1}{2}} (-2\beta_x J_x \sigma_y^4)^{k}\left( \sqrt{2\beta_y J_y}\sigma_x^2 \right)^{n-2k-1} + n\binom{2l+1}{l} \binom{u+n-2l-1}{\frac{u+n-2l-1}{2}}  ~\times \\ 
	&\left( \sqrt{2\beta_y J_y}\sigma_x^2 \right)^{n-1}  \Bigg) + \frac{\mathrm{i} \widetilde{c_u w_m} \sqrt{2\beta_x J_x}}{2^{u+m}\Delta_x^m} \Bigg( \sum_{p=1}^{\frac{m-3}{2}}\frac{m}{2p+1}   \binom{m-1}{2p}\binom{u+m-2l-2p-1}{\frac{u+m-2l-2p-1}{2}}\binom{2l+2p+1}{l+p} ~\times \\
	&\left( -\sqrt{2\beta_y J_y} \right)^{m-2p-1} (-2\beta_x J_x)^{p} +  m\binom{u+m-2l-1}{\frac{u+m-2l-1}{2}}\binom{2l+1}{l} \left( \sqrt{2\beta_y J_y} \right)^{m-1} \Bigg) \Bigg) + \sum_{j=1}^{u}\binom{u}{2j-1} ~\times \\ 
	&\left( \sqrt{2\beta_x J_x} \right)^{2j-1} \left( \mathrm{i}\sqrt{2\beta_y J_y} \right)^{u-2j+1}  \Biggg( \frac{\widetilde{c_u g_n}}{2^{u+n}(\sigma_x\sigma_y\Delta_x)^n}  \binom{u+n-2j+1}{\frac{u+n-2j+1}{2}}\binom{2j-1}{j-1} \left( \sqrt{2\beta_y J_y}\sigma_x^2 \right)^n ~+ \\
	&\frac{\widetilde{c_u w_m}}{2^{u+m}\Delta_x^m} \binom{2j-1}{j-1} \binom{u+m-2j+1}{\frac{u+m-2j+1}{2}} \left( -\sqrt{2\beta_y J_y} \right)^m + \frac{\widetilde{c_u g_n} \sigma_x^2 \sqrt{2\beta_y J_y}}{2^{u+n}(\sigma_x\sigma_y\Delta_x)^n} \Bigg( \sum_{k=1}^{\frac{n-3}{2}}\frac{n}{n-2k} \binom{n-1}{2k} ~\times \\ 
	&\binom{u+n-2j-2k+1}{\frac{u+n-2j-2k+1}{2}}\binom{2j+2k-1}{j+k-1}  \left( \sqrt{2\beta_y J_y}\sigma_x^2 \right)^{n-2k-1} (-2\beta_x J_x \sigma_y^4)^k +  n \binom{u-2j+2}{\frac{u-2j+2}{2}}\binom{n+2j-2}{\frac{n+2j-3}{2}} ~\times  \\
	&\left( -\mathrm{i}  \sigma_y^2 \sqrt{2\beta_x J_x}\right)^{n-1} \Bigg) - \frac{\widetilde{c_u w_m}\sqrt{2\beta_y J_y}}{2^{u+m}\Delta_x^m} \Biggg( \sum_{p=1}^{\frac{m-3}{2}} \frac{m}{m-2p} \binom{m-1}{2p} \binom{u+m-2j-2p+1}{\frac{u+m-2j-2p+1}{2}} ~\times \\ 
	& \binom{2j+2p-1}{j+p-1} (-2\beta_x J_x)^p \left( -\sqrt{2\beta_y J_y} \right)^{m-2p-1} +  m \binom{u-2j+2}{\frac{u-2j+2}{2}} \binom{m+2j-2}{\frac{m+2j-3}{2}} \left( \mathrm{i}\sqrt{2\beta_x J_x} \right)^{m-1}  \Biggg)  \Biggg)    \Biggg] 
	\end{split}
	\end{equation}
	\begin{equation}
	\begin{split}
	\label{tsa_fxodev_x}
	&\mathcal{T}1_x^{(u,n,m)}= \sqrt{\frac{\beta_x}{2J_x}}\Bigg[  \frac{\widehat{c_u g_n}}{2^{u+n}(\sigma_x\sigma_y\Delta_x)^n} \Bigg( \sum_{k=0}^{\frac{n}{2}} \binom{n}{2k}\left( -2\beta_x J_x \sigma_y^4\right)^k \left( \sqrt{2\beta_y J_y}\sigma_x^2\right)^{n-2k} \Bigg( \binom{n-2k}{\frac{n-2k}{2}}\binom{u+2k}{\frac{u+2k-1}{2}} ~\times \\
	&\left( \sqrt{2\beta_x J_x}\right)^u + \sqrt{2\beta_x J_x} \Bigg( \sum_{l=1}^{\frac{u-3}{2}} \frac{u}{2l+1}\binom{u-1}{2l}\binom{u+n-2l-2k-1}{\frac{u+n-2l-2k-1}{2}} \binom{2l+2k+1}{l+k} \left( 2\beta_x J_x \right)^l ~\times \\
	&\left( \mathrm{i}\sqrt{2\beta_y J_y} \right)^{u-2l-1} +  u\binom{2k+1}{k} \binom{u+n-2k-1}{\frac{u+n-2k-1}{2}}\left( \mathrm{i}\sqrt{2\beta_yJ_y}\right)^{u-1}  \Bigg) \Bigg) + \sum_{j=0}^{n} \binom{n}{2j-1} \left( -\mathrm{i}\sqrt{2\beta_x J_x}\sigma_y^2 \right)^{2j-1} ~\times \\
	&\left( \sqrt{2\beta_y J_y}\sigma_x^2 \right)^{n-2j+1} \Bigg(  \binom{2j-1}{j-1} \binom{u+n-2j+1}{\frac{u+n-2j+1}{2}} \left( \mathrm{i} \sqrt{2\beta_y J_y}\right)^u + \mathrm{i} \sqrt{2\beta_y J_y} \Bigg( \sum_{l=1}^{\frac{u-3}{2}}\frac{u}{u-2l}\binom{u-1}{2l} ~\times \\ 
	&\binom{u+n-2l-2j+1}{\frac{u+n-2l-2j+1}{2}} \binom{2l+2j-1}{l+j-1}\left( 2\beta_x J_x \right)^l \left( \mathrm{i}\sqrt{2\beta_y J_y} \right)^{u-2l-1} + u\binom{n-2j+2}{\frac{n-2j+2}{2}} \binom{u+2j-2}{\frac{u+2j-3}{2}} ~\times \\
	&\left( \sqrt{2\beta_x J_x}\right)^{u-1}  \Bigg)  \Bigg)  \Bigg) + \frac{\widehat{c_u w_m}}{2^{u+m}\Delta_x^m} \Biggg( \sum_{p=0}^{\dfrac{m}{2}}\binom{m}{2p} \left( -2\beta_x J_x \right)^p \left( -\sqrt{2\beta_y J_y}\right)^{m-2p}  \Biggg( \binom{m-2p}{\frac{m-2p}{2}}\binom{u+2p}{\frac{u+2p-1}{2}} ~\times \\
	&\left( \sqrt{2\beta_x J_x} \right)^u + \sqrt{2\beta_x J_x} \Bigg( \sum_{l=1}^{\frac{u-3}{2}}\frac{u}{2l+1} \binom{u-1}{2l} \binom{2l+2p+1}{l+p} \binom{u+m-2l-2p-1}{\frac{u+m-2l-2p-1}{2}} \left( 2\beta_x J_x \right)^l ~\times \\ 
	& \left( \mathrm{i} \sqrt{2\beta_y J_y} \right)^{u-2l-1} + u\binom{2p+1}{p} \binom{u+m-2p-1}{\frac{u+m-2p-1}{2}} \left( \mathrm{i} \sqrt{2\beta_y J_y} \right)^{u-1}  \Bigg) \Biggg) + \sum_{h=0}^{m}\binom{m}{2h-1} \left( \mathrm{i}\sqrt{2\beta_x J_x} \right)^{2h-1} ~\times \\ 
	&\left( -\sqrt{2\beta_y J_y} \right)^{m-2h+1}  \Biggg( \binom{u+m-2h+1}{\frac{u+m-2h+1}{2}}\binom{2h-1}{h-1}  \left( \mathrm{i}\sqrt{2\beta_y J_y} \right)^u + \mathrm{i}\sqrt{2\beta_y J_y} \Biggg( \sum_{l=1}^{\frac{u-3}{2}}\frac{u}{u-2l} \binom{u-1}{2l} ~\times \\
	&\binom{2l+2h-1}{l+h-1} \binom{u+m-2l-2h+1}{\frac{u+m-2l-2h+1}{2}} \left( 2\beta_x J_x \right)^l \left( \mathrm{i} \sqrt{2\beta_y J_y} \right)^{u-2l-1} ~+ \\
	&u\binom{m-2h+2}{\frac{m-2h+2}{2}}  \binom{u+2h-2}{\frac{u+2h-3}{2}} \left( \sqrt{2\beta_x J_x} \right)^{u-1}  \Biggg) \Biggg) \Biggg) \Biggg]
	\end{split}
	\end{equation}
	\begin{equation}
	\begin{split}
	\label{tsa_fxevod_y}
	&\mathcal{T}_y^{(u,n,m)}= \sqrt{\frac{\beta_y}{2J_y}}\Biggg[ \sum_{l=0}^{\frac{u}{2}}\binom{u}{2l}(-2\beta_y J_y)^{l}  \left(\sqrt{2\beta_x J_x} \right)^{u-2l} \Bigg( \frac{\widetilde{c_u g_n}}{2^{u+n}(\sigma_x\sigma_y\Delta_x)^n} \binom{u-2l}{\frac{u-2l}{2}}\binom{n+2l}{\frac{n+2l-1}{2}} ~\times \\
	&\left(\sqrt{2\beta_y J_y}\sigma_x^2 \right)^n  + \frac{\widetilde{c_u w_m}\left( -\sqrt{2\beta_y J_y} \right)^m}{2^{u+m}\Delta_x^m}\binom{u-2l}{\frac{u-2l}{2}} \binom{m+2l}{\frac{m+2l-1}{2}} +\frac{\widetilde{c_u g_n}\sqrt{2\beta_y J_y}\sigma_x^2 }{2^{u+n}(\sigma_x\sigma_y\Delta_x)^n}  \Bigg( \sum_{k=1}^{\frac{n-3}{2}}\frac{n}{2k+1}\binom{n-1}{2k} ~\times \\
	&\binom{2l+2k+1}{l+k} \binom{u+n-2l-2k-1}{\frac{u+n-2l-2k-1}{2}} (2\beta_y J_y \sigma_x^4)^{k}\left( -\mathrm{i} \sqrt{2\beta_x J_x}\sigma_y^2 \right)^{n-2k-1} + n \binom{2l+1}{l} \binom{u+n-2l-1}{\frac{u+n-2l-1}{2}} ~\times \\ 
	&\left( -\mathrm{i} \sqrt{2\beta_x J_x}\sigma_y^2 \right)^{n-1}  \Bigg) - \frac{\widetilde{c_u w_m} \sqrt{2\beta_y J_y}}{2^{u+m}\Delta_x^m} \Bigg( \sum_{p=1}^{\frac{m-3}{2}}\frac{m}{2p+1}   \binom{m-1}{2p}\binom{u+m-2l-2p-1}{\frac{u+m-2l-2p-1}{2}}\binom{2l+2p+1}{l+p} ~\times \\
	&\left( \mathrm{i} \sqrt{2\beta_x J_x} \right)^{m-2p-1} (2\beta_y J_y)^{p} ~+ m \binom{u+m-2l-1}{\frac{u+m-2l-1}{2}}\binom{2l+1}{l} \left( \mathrm{i} \sqrt{2\beta_x J_x} \right)^{m-1} \Bigg) \Bigg) + \sum_{j=1}^{u}\binom{u}{2j-1}  ~\times \\ 
	&\left( \mathrm{i} \sqrt{2\beta_y J_y} \right)^{2j-1} \left(\sqrt{2\beta_x J_x} \right)^{u-2j+1} \Biggg( \frac{\widetilde{c_u g_n}}{2^{u+n}(\sigma_x\sigma_y\Delta_x)^n} \binom{u+n-2j+1}{\frac{u+n-2j+1}{2}}\binom{2j-1}{j-1} \left( -\mathrm{i} \sqrt{2\beta_x J_x}\sigma_x^2 \right)^n ~+ \\ 
	&\frac{\widetilde{c_u w_m}}{2^{u+m}\Delta_x^m} \binom{2j-1}{j-1} \binom{u+m-2j+1}{\frac{u+m-2j+1}{2}} \left( \mathrm{i} \sqrt{2\beta_x J_x} \right)^m - \frac{\mathrm{i} \widetilde{c_u g_n} \sigma_y^2 \sqrt{2\beta_x J_x}}{2^{u+n}(\sigma_x\sigma_y\Delta_x)^n} \Biggg( \sum_{k=1}^{\frac{n-3}{2}}\frac{n}{n-2k}   \binom{n-1}{2k} ~\times \\ 
	&\binom{u+n-2j-2k+1}{\frac{u+n-2j-2k+1}{2}}\binom{2j+2k-1}{j+k-1} \left( -\mathrm{i} \sqrt{2\beta_x J_x}\sigma_y^2 \right)^{n-2k-1} (2\beta_y J_y \sigma_x^4)^k + n \binom{u-2j+2}{\frac{u-2j+2}{2}}\binom{n+2j-2}{\frac{n+2j-3}{2}} ~\times \\
	& \left( \sigma_x^2 \sqrt{2\beta_y J_y}\right)^{n-1} \Biggg) +   \frac{\mathrm{i}\widetilde{c_u w_m} \sqrt{2\beta_x J_x}}{2^{u+m}\Delta_x^m} \Bigg( \sum_{p=1}^{\frac{m-3}{2}} \frac{m}{m-2p} \binom{m-1}{2p} \binom{u+m-2j-2p+1}{\frac{u+m-2j-2p+1}{2}} ~\times \\ 
	& \binom{2j+2p-1}{j+p-1} (2\beta_y J_y)^p \left( \mathrm{i} \sqrt{2\beta_x J_x} \right)^{m-2p-1} +  m \binom{u-2j+2}{\frac{u-2j+2}{2}} \binom{m+2j-2}{\frac{m+2j-3}{2}} \left( -\sqrt{2\beta_y J_y} \right)^{m-1}  \Bigg)  \Biggg)    \Biggg] 
	\end{split}
	\end{equation}
	\begin{equation}
	\begin{split}
	\label{tsa_fxodev_y}
	&\mathcal{T}1_y^{(u,n,m)}= \sqrt{\frac{\beta_y}{2J_y}}\Biggg[  \frac{\widehat{c_u g_n}}{2^{u+n}(\sigma_x\sigma_y\Delta_x)^n} \Biggg( \sum_{k=0}^{\frac{n}{2}} \binom{n}{2k} \left( 2\beta_y J_y \sigma_x^4 \right)^k \left( -\mathrm{i}\sqrt{2\beta_x J_x}\sigma_y^2\right)^{n-2k} \Bigg( \binom{n-2k}{\frac{n-2k}{2}}\binom{u+2k}{\frac{u+2k-1}{2}} \times \\
	&\left( \sqrt{2\beta_x J_x}\right)^u + \mathrm{i}\sqrt{2\beta_y J_y} \Bigg( \sum_{l=1}^{\frac{u-3}{2}}  \frac{u}{2l+1}\binom{u-1}{2l}\binom{u+n-2l-2k-1}{\frac{u+n-2l-2k-1}{2}} \binom{2l+2k+1}{l+k} \left( -2\beta_y J_y\right)^l ~\times \\
	&\left( \sqrt{ 2\beta_x J_x} \right)^{u-2l-1} + u\binom{2k+1}{k} \binom{u+n-2k-1}{\frac{u+n-2k-1}{2}}\left( \sqrt{2\beta_x J_x}\right)^{u-1}  \Bigg) \Bigg) + \sum_{j=0}^{n} \binom{n}{2j-1} \left( \sqrt{2\beta_y J_y}\sigma_x^2 \right)^{2j-1} ~\times \\
	&\left( -\mathrm{i}\sqrt{2\beta_x J_x}\sigma_y^2 \right)^{n-2j+1}  \Biggg(  \binom{2j-1}{j-1} \binom{u+n-2j+1}{\frac{u+n-2j+1}{2}}  \left( \sqrt{2\beta_x J_x}\right)^u + \sqrt{2\beta_x J_x} \Bigg( \sum_{l=1}^{\frac{u-3}{2}}\frac{u}{u-2l}\binom{u-1}{2l}~ \times \\ 
	&\binom{u+n-2l-2j+1}{\frac{u+n-2l-2j+1}{2}}  \binom{2l+2j-1}{l+j-1} \left( -2\beta_y J_y \right)^l\left( \sqrt{2\beta_x J_x} \right)^{u-2l-1} + u\binom{n-2j+2}{\frac{n-2j+2}{2}} \binom{u+2j-2}{\frac{u+2j-3}{2}} ~\times \\
	&\left( \mathrm{i}\sqrt{2\beta_y J_y}\right)^{u-1}  \Bigg)  \Biggg)  \Biggg) + \frac{\widehat{c_u w_m}}{2^{u+m}\Delta_x^m} \Biggg( \sum_{p=0}^{\dfrac{m}{2}}\binom{m}{2p}\left( 2\beta_y J_y \right)^p \left( \mathrm{i}\sqrt{2\beta_y J_y}\right)^{m-2p} \Bigg( \binom{m-2p}{\frac{m-2p}{2}}\binom{u+2p}{\frac{u+2p-1}{2}} ~\times \\
	&\left( \mathrm{i}\sqrt{2\beta_y J_y} \right)^u  + \mathrm{i}\sqrt{2\beta_y J_y} \Bigg( \sum_{l=1}^{\frac{u-3}{2}}\frac{u}{2l+1} \binom{u-1}{2l} \binom{2l+2p+1}{l+p} \binom{u+m-2l-2p-1}{\frac{u+m-2l-2p-1}{2}} \left( -2\beta_y J_y \right)^l ~\times \\ 
	&\left(  \sqrt{2\beta_x J_x} \right)^{u-2l-1} + u\binom{2p+1}{p} \binom{u+m-2p-1}{\frac{u+m-2p-1}{2}} \left( \sqrt{2\beta_x J_x} \right)^{u-1}  \Bigg) \Bigg) + \sum_{h=0}^{m}\binom{m}{2h-1} \left( -\sqrt{2\beta_y J_y} \right)^{2h-1} ~\times  \\ 
	& \left( \mathrm{i}\sqrt{2\beta_x J_x} \right)^{m-2h+1}   \Biggg( \binom{u+m-2h+1}{\frac{u+m-2h+1}{2}}\binom{2h-1}{h-1} \left( \sqrt{2\beta_x J_x} \right)^u + \sqrt{2\beta_x J_x} \Biggg( \sum_{l=1}^{\frac{u-3}{2}}\frac{u}{u-2l} \binom{u-1}{2l} ~\times \\
	&\binom{2l+2h-1}{l+h-1} \binom{u+m-2l-2h+1}{\frac{u+m-2l-2h+1}{2}} \left( -2\beta_y J_y \right)^l \left( \mathrm{i} \sqrt{2\beta_x J_x} \right)^{u-2l-1}  + u\binom{m-2h+2}{\frac{m-2h+2}{2}} ~\times \\ 
	&  \binom{u+2h-2}{\frac{u+2h-3}{2}} \left( \mathrm{i}\sqrt{2\beta_y J_y} \right)^{u-1}  \Biggg) \Biggg) \Biggg) \Biggg]
	\end{split}
	\end{equation}
	\begin{equation}
		\widetilde{c_u g_n}=Re[c_u g_n]
	\end{equation}
	\begin{equation}
		\widetilde{c_u w_m}=Re[c_u w_m]
	\end{equation}
	\begin{equation}
		\widehat{c_u g_n}=\mathrm{i}  Im[c_u g_n]
	\end{equation}
	\begin{equation}
		\widehat{c_u w_m}=\mathrm{i}  Im[c_u w_m]
	\end{equation}
\end{subequations}
}
\normalsize
where the $c_u$, $g_n$ and $w_m$ are given by Eqs.~(\ref{coef_elxy}).

\subsection{TSA$_\chi$ driven by elliptical bunches with $\sigma_y>\sigma_x$}
The TSA$_\chi$ from a BBLR kick that is generated by an elliptical strong bunch with $\sigma_y>\sigma_x$ is described by:
\begin{subequations}
	\begin{equation}
	\begin{split}
	\text{TSA}_x=& -\frac{q~\beta_{ls}}{2\pi ~c ~P_0} \mathcal{L}_y \Bigg[\sum_{u=0,2,4,\dots}^{\infty}\sum_{n=1,3,5,\dots}^{\infty}\sum_{m=1,3,5,\dots}^{\infty}\mathcal{T}2_x^{(u,n,m)}~+ \\ 
	&\sum_{u=1,3,5,\dots}^{\infty}\sum_{n=0,2,4,\dots}^{\infty}\sum_{m=0,2,4,\dots}^{\infty}\mathcal{T}3_x^{(u,n,m)} \Bigg]
	\end{split}
	\end{equation}
	\begin{equation}
	\begin{split}
	\text{TSA}_y=& \mathrm{i}  \frac{q~\beta_{ls}}{2\pi ~c ~P_0} \mathcal{L}_y \Bigg[\sum_{u=0,2,4,\dots}^{\infty}\sum_{n=1,3,5,\dots}^{\infty}\sum_{m=1,3,5,\dots}^{\infty}\mathcal{T}2_y^{(u,n,m)}~ + \\  &\sum_{u=1,3,5,\dots}^{\infty}\sum_{n=0,2,4,\dots}^{\infty}\sum_{m=0,2,4,\dots}^{\infty}\mathcal{T}3_y^{(u,n,m)} \Bigg]
	\end{split}
	\end{equation}
\end{subequations}
where the $\mathcal{L}_y$ is equal to $\mathcal{L}$ (Eq.~(\ref{field_L})) when $\sigma_y>\sigma_x$. The $\mathcal{T}2_x^{(u,n,m)}$, $\mathcal{T}3_x^{(u,n,m)}$, $\mathcal{T}2_y^{(u,n,m)}$ and $\mathcal{T}3_y^{(u,n,m)}$ can be obtained from the $\mathcal{T}_y^{(u,n,m)}$ (Eq.~(\ref{tsa_fxevod_y})), $\mathcal{T}1_y^{(u,n,m)}$ (Eq.~(\ref{tsa_fxodev_y})), $\mathcal{T}_x^{(u,n,m)}$ (Eq.~(\ref{tsa_fxevod_x})) and $\mathcal{T}1_x^{(u,n,m)}$ (Eq.~(\ref{tsa_fxodev_x})) respectively provided that $x$ is replaced by $y$ and $y$ by $x$. The $c_u$, $g_n$ and $w_m$ are given by Eqs.~(\ref{coef_elyx_first}-\ref{coef_elyx_last}).

\subsection{TSA$_\chi$ driven by round bunches with $\mathbf{\sigma_x=\sigma_y}$} 
The BBLR kick from a round strong bunch ($\sigma_x=\sigma_y=\sigma$) generate TSA$_\chi$ that is given by the following equations:
\begin{subequations}
	\begin{equation}
	\text{TSA}_x= -\frac{q~\beta_{ls}}{2\pi ~c ~P_0} \mathcal{L}_r \left\langle f_r \mathtt{Cos}\left[ \phi_x\right]\left( \beta_x \mathtt{Cos}\left[ \phi_x\right]+ \widetilde{x}\sqrt{\frac{\beta_x}{2J_x}} \right) \right\rangle_\phi\
	\end{equation}
	\begin{equation}
	\text{TSA}_y= -\frac{q~\beta_{ls}}{2\pi ~c ~P_0} \mathcal{L}_r \left\langle f_r \mathtt{Cos}\left[ \phi_y\right] \left( \beta_y \mathtt{Cos}\left[ \phi_y\right]+ \widetilde{y}\sqrt{\frac{\beta_y}{2J_y}} \right) \right\rangle_\phi\ .
	\end{equation}
\end{subequations}
The $\mathcal{L}_r$ is equal to $\mathcal{L}$ (Eq.~(\ref{field_L})) when $\sigma_x=\sigma_y$ and the $f_r$ is given by the Eq.~(\ref{fr}). Integrating over the angles with the help of the Eqs.~(\ref{trig_red}), the TSA$_\chi$ at the $x$ and $y$ plane is defined by:
\begin{subequations}
	\begin{equation}
	\text{TSA}_x=-\frac{q~\beta_{ls}}{2\pi ~c ~P_0} ~\mathcal{L}_r\sum_{u=0}^{\infty}\mathcal{T}4_x^{(u)}
	\end{equation}
	\begin{equation}
	\text{TSA}_y=-\frac{q~\beta_{ls}}{2\pi ~c ~P_0} ~\mathcal{L}_r\sum_{u=0}^{\infty}\mathcal{T}4_y^{(u)}
	\end{equation}
	\begin{equation}
	\begin{split}
	\mathcal{T}4_x^{(u)}=&\frac{(-1)^u}{2^{u+1}~(u+1)!~\sigma^{2(u+1)}} \Bigg[ \sum_{l=0}^{u}\sum_{m=0}^{u-l} \frac{1}{4^{m}} \binom{u}{l}\binom{2(u-l)}{2m}\binom{2m}{m} (\sqrt{2\beta_y J_y})^{2m} ~\widetilde{y}^{2(u-l-m)} ~\times \\
	& \Bigg( \sum_{n_1=0}^{l}\frac{1}{4^{n_1+1}}\binom{2l}{2n_1}\binom{2(n_1+1)}{n_1+1} \beta_x(\sqrt{2\beta_xJ_x})^{2n_1} ~\widetilde{x}^{2(l-n_1)} ~+ \\ 
	& \sum_{n_2=0}^{l-1}\frac{1}{4^{n_2+1}}\binom{2l}{2n_2+1}\binom{2(n_2+1)}{n_2+1} \sqrt{\frac{\beta_x}{2 J_x}} (\sqrt{2\beta_xJ_x})^{2n_2+1} ~\widetilde{x}^{2(l-n_2)} \Bigg)  \Bigg] 
	\end{split}
	\end{equation}
	\begin{equation}
	\begin{split}
	\mathcal{T}4_y^{(u)}=&\frac{(-1)^u}{2^{u+1}~(u+1)!~\sigma^{2(u+1)}} \Bigg[ \sum_{l=0}^{u}\sum_{n=0}^l \frac{1}{4^{n}} \binom{u}{l}\binom{2l}{2n}\binom{2n}{n} (\sqrt{2\beta_x J_x})^{2n} ~\widetilde{x}^{2(l-n)} ~\times \\
	& \Bigg( \sum_{m_1=0}^{u-l}\frac{1}{4^{m_1+1}}\binom{2(u-l)}{2m_1}\binom{2(m_1+1)}{m_1+1} \beta_y(\sqrt{2\beta_yJ_y})^{2m_1} ~\widetilde{y}^{2(u-l-m_1)} ~+ \\ 
	& \sum_{m_2=0}^{u-l-1}\frac{1}{4^{m_2+1}}\binom{2(u-l)}{2m_2+1}\binom{2(m_2+1)}{m_2+1} \sqrt{\frac{\beta_y}{2 J_y}} (\sqrt{2\beta_yJ_y})^{2m_2+1} ~\widetilde{y}^{2(u-l-m_2)} \Bigg)  \Bigg] .
	\end{split}
	\end{equation}
\end{subequations}
\subsection{TSA$_\chi$ driven by a DC wire}
For a DC wire, the Eqs.~(\ref{tsa_bef}) take the following form:
\begin{subequations}
    \label{tsa_wire_in}
	\begin{equation}
	\text{TSA}_x= \frac{1}{2\pi}\sqrt{\frac{\beta_x}{2J_x}} \frac{q}{P_0} \mathcal{Z}_w Im\left[  \left\langle f_w~\mathtt{Cos}\left[ \phi_x\right] \right\rangle_\phi\right] 
	\end{equation}
	\begin{equation}
	\text{TSA}_y= -\frac{1}{2\pi}\sqrt{\frac{\beta_y}{2J_y}} \frac{q}{P_0} \mathcal{Z}_w Re\left[ \left\langle f_w~\mathtt{Cos}\left[ \phi_y\right] \right\rangle_\phi\right], 
	\end{equation}
\end{subequations}
where the $\mathcal{L}_w$ is given by the Eq.~(\ref{coef_w}) and the $f_w$ by the Eqs.~(\ref{fw}) and (\ref{fwu}). Only the odd powers of $u$ contribute to the TSA$_\chi$ therefore, the Eqs.~(\ref{tsa_wire_in}) after the integration over the angles conclude to the following ones:
\begin{subequations}
	\label{tsa_wire}
	\begin{equation}
	\text{TSA}_x= \frac{1}{2\pi}\sqrt{\frac{\beta_x}{2J_x}} \frac{q}{P_0}~\mathcal{L}_w\sum_{u=1,3,5,\dots}^{\infty}\mathcal{T}5_x^{(u)}
	\end{equation}
	\begin{equation}
	\text{TSA}_y= -\frac{1}{2\pi}\sqrt{\frac{\beta_y}{2J_y}} \frac{q}{P_0}~\mathcal{L}_w\sum_{u=1,3,5,\dots}^{\infty}\mathcal{T}5_y^{(u)}
	\end{equation}
	\begin{equation}
	\begin{split}
	\mathcal{T}5_x^{(u)}=&\widetilde{c_u}\Bigg(\frac{1}{2^{u+1}}\sqrt{\frac{\beta_x}{2 J_x}}\left(\sqrt{2 \beta_x J_x} \right)^u \binom{u+1}{\frac{u+1}{2}} + \frac{u}{2^u} \beta_x \left(\mathrm{i}  \sqrt{2 \beta_y J_y} \right)^{u-1}\binom{u-1}{\frac{u-1}{2}} ~+ \\
	&\sqrt{\frac{\beta_x}{2 J_x}}\sum_{l=1}^{\frac{u-3}{2}}\frac{u}{2l-1} \frac{1}{4^{l+1}}\frac{1}{2^{u-2l-1}} \left(\sqrt{2 \beta_x J_x} \right)^{2l+1} \left(  \mathrm{i}\sqrt{2 \beta_y J_y} \right)^{u-2l-1} ~\times \\
	&\binom{u-1}{2l}\binom{2l+2}{l+1}\binom{u-2l-1}{\frac{u-2l-1}{2}} \Bigg)
	\end{split}
	\end{equation}
	\begin{equation}
	\begin{split}
	\mathcal{T}5_y^{(u)}=&\widehat{c_u}\Bigg(\frac{1}{2^{u+1}}\sqrt{\frac{\beta_y}{2 J_y}}\left(  \mathrm{i}\sqrt{2 \beta_y J_y} \right)^u \binom{u+1}{\frac{u+1}{2}} + \frac{ \mathrm{i} u}{2^u} \beta_y \left( \sqrt{2 \beta_x J_x} \right)^{u-1}\binom{u-1}{\frac{u-1}{2}} ~+ \\
	&\sqrt{\frac{\beta_y}{2 J_y}}\sum_{l=1}^{\frac{u-3}{2}}\frac{u}{u-2l} \frac{1}{4^l}\frac{1}{2^{u-2l+1}} \left( \sqrt{2 \beta_x J_x} \right)^{2l} \left( \mathrm{i}\sqrt{2 \beta_y J_y} \right)^{u-2l} \binom{u-1}{2l}\binom{2l}{l}\binom{u-2l+1}{\frac{u-2l+1}{2}}\Bigg) 
	\end{split}
	\end{equation}
	\begin{equation}
		\widetilde{c_u}=Re[c_u]=\frac{1}{\sum_{n=0}^{\frac{u+1}{2}}~\binom{u+1}{2n}~ x_{wb}^{2n}~\left(\mathrm{i} y_{wb}\right)^{u+1-2n}}
	\end{equation}
	\begin{equation}
		\widehat{c_u}=\mathrm{i} Im[c_u]=\frac{1}{\sum_{n=0}^{\frac{u+1}{2}-1}~\binom{u+1}{2n+1}~ x_{wb}^{2n+1}~\left(\mathrm{i} y_{wb}\right)^{u-2n}}.
	\end{equation}
\end{subequations}

\addcontentsline{toc}{section}{References}
\bibliographystyle{amsplain}
\bibliography{my_ref}

\providecommand{\bysame}{\leavevmode\hbox to3em{\hrulefill}\thinspace}
\providecommand{\MR}{\relax\ifhmode\unskip\space\fi MR }
\providecommand{\MRhref}[2]{%
  \href{http://www.ams.org/mathscinet-getitem?mr=#1}{#2}
}
\providecommand{\href}[2]{#2}
\begin{thebibliography}{1}

\bibitem{baer}
M~Bassetti and George~A Erskine, \emph{{Closed expression for the electrical
  field of a two-dimensional Gaussian charge}}, Tech. Report CERN-ISR-TH-80-06.
  ISR-TH-80-06, CERN, Geneva, 1980.

\bibitem{beng}
Johan Bengtsson, \emph{{Non-linear transverse dynamics for storage rings with
  applications to the low-energy antiproton ring (LEAR) at CERN}}, Ph.D.
  thesis, Lund U., Geneva, 1998.

\bibitem{gold}
Herbert Goldstein, Charles Poole, and John Safko, \emph{{Classical Mechanics;
  3rd ed.}}, Addison-Wesley, San Francisco, CA, 2002.

\bibitem{guig}
Gilbert Guignard, \emph{A general treatment of resonances in accelerators},
  CERN, CERN, 1978, CERN, Geneva, 1977 - 1978, p.~72 p.

\bibitem{hira}
K~Hirata, Herbert~W Moshammer, and F~Ruggiero, \emph{{A symplectic beam-beam
  interaction with energy change}}, Part. Accel. \textbf{40} (1992),
  no.~KEK-92-117, 205--228. 25 p.

\bibitem{jack}
John~David Jackson, \emph{Classical electrodynamics}, 3rd ed. ed., Wiley, New
  York, {NY}, 1999.

\end{thebibliography}

\end{document}